\documentclass[12pt]{article}
\usepackage{natbib}
\pdfoutput=1 
\usepackage{amsmath,amsfonts,amssymb}
\usepackage[english]{babel}
\usepackage{graphicx}
\usepackage{color}
\usepackage{xtab}
\newcommand{\Third}{{\textstyle{\frac{3}{2}}}}
\newcommand{\half}{{\textstyle{\frac{1}{2}}}}
\newcommand{\be}{\begin{equation}}
\newcommand{\ee}{\end{equation}}
\newcommand{\bea}{\begin{eqnarray}}
\newcommand{\eea}{\end{eqnarray}}

\begin{document}


\pagestyle{empty}

\bigskip

\bigskip

\begin{center}

\vfill 
 
 \vspace{30mm}

{\Large \textbf{\textsf{Correlations in  Usage Frequencies and Shannon Entropy for Codons}}}

\vspace{10mm}

{\large D. Cocurullo $^{a}$, A. Sciarrino$^{ab}$}

\vspace{10mm}

\emph{$^a$ Dipartimento di Scienze Fisiche, Universit{\`a} di Napoli
``Federico II''}

\emph{$^b$ I.N.F.N., Sezione di Napoli}

\emph{Complesso Universitario di Monte S. Angelo}

\emph{Via Cinthia, I-80126 Napoli, Italy}

\vspace{12mm}

\end{center}

\vspace{7mm}

\begin{abstract}
 The usage frequencies for  codons belonging to quartets  are analized, over the whole exonic region, for 92 biological species.  Correlation is put into evidence, between the usage frequencies of synonymous codons with third nucleotide A and C and between the usage frequencies of non synonymous codons, belonging to suitable subsets of the quartets, with the same third nucleotide.   A correlation is pointed out between   amino acids belonging to subsets of the set encoded by quartets of codons.  It is remarked that the computed Shannon entropy for quartets is weakly dependent on the biological species. The observed correlations well fit in the mathematical scheme of the crystal basis model of the genetic code.
\end{abstract}

\vfill
\bigskip

\vfill

\leftline{DSF-Th-2/08-v2}

\noindent
\emph{E-mail:} \texttt{sciarrino@na.infn.it}

\newpage

\pagestyle{plain}
\setcounter{page}{1} 

\section{Introduction}

The genetic information in DNA is stored in sequences built up from four
bases (nucleotides)\footnote{In the paper we denote the nucleotides and the amino acids   by their initial letter or by the abbreviation of their name, according to the standard convention} $C$, $T$, $G$, $A$ (in mRNA, which plays a key role in
the construction of proteins,  $T$ is replaced by $U$).  The
proteins are made up from 20 different amino-acids (a.a.). The ``quantum'' of
genetic information is constituted by an ordered triplet of nucleotides
(codon). There are therefore 64 possible codons, which encode 20
amino-acids, plus the three signals (in the eukaryotic code) of the termination of the biosynthesis
process (stop codons). It follows that the genetic code, i.e. the correspondence between
codons and amino-acids, is degenerate.
Degeneracy refers to the fact that almost
all the  a.a. are encoded by multiple codons (called synonymous codons), see the Genetic Code Table 1.

 Degeneracy is found
primarily in the third position of the codon, i.e. the nucleotide in the
third position can change without changing the corresponding a.a.
 The currently available data show that some codons are used much more
frequently than others to encode a particular amino-acid, i.e.
there is a  ``codon bias".  It is currently believed that a non-uniform usage of
synonymous codons is a widespread phenomenon and it is experimentally
observed that the pattern of codon usage varies between species and even
between tissues within a species; see refs. \citep{dumou99,  kana2001}, which
contain a large number of references to the original works on the subject.
The main reasons for the codon usage biases are believed to be: the
mutational biases, the translation efficiency, the natural selection and
the abundance of specific anticodons in the tRNA.
The aim of this
paper is not to discuss or to compare the different proposed explanations, but to search for 
  a possible general pattern of the bias.

Most of the studies of the codon usage frequencies has addressed to the
analysis of the relative abundance of a specified codon in different genes
of the same biological species or in the comparison of the relative
abundance in the same gene for different biological species. Little
attention has been paid to analyze the codon usage frequency summed over
the whole available sequences to infer global correlations between
different biological species. Indeed, in \citep{KFL}, analysing a large
sample of species, a correlation between the GC content and the codon usage
has been pointed out and explained on the basis of a mutational model at
the nucleotide level. 

\begin{table}[htbp]
\footnotesize
\caption{The eukaryotic or standard code code. }
\label{tablegc}
\footnotesize
\begin{center}
\begin{tabular}{|cc||cc||cc||cc|}
\hline
codon & a.a.  &  codon & a.a. &   codon & a.a. &   codon & a.a.  \\
\hline
CCC & Pro P & UCC & Ser S &  GCC & Ala A &   ACC & Thr T   \\
CCU & Pro P & UCU & Ser S & GCU & Ala A &    ACU & Thr T   \\
CCG & Pro P &  UCG & Ser S & GCG & Ala A & ACG & Thr T   \\
CCA & Pro P &  UCA & Ser S & GCA & Ala A &  ACA & Thr T \\[1mm] \hline
CUC & Leu L &  UUC & Phe F & GUC & Val V &   AUC & Ile I   \\
CUU & Leu L &  UUU & Phe F & GUU & Val V  & AUU & Ile I  \\
CUG & Leu L &   UUG & Leu L &  GUG & Val V &  AUG & Met M \\
CUA & Leu L &   UUA & Leu L & GUA & Val V &  AUA & Ile I  \\[1mm] \hline
CGC & Arg R &   UGC & Cys C & GGC & Gly G &   AGC & Ser S  \\
CGU & Arg R &  UGU & Cys C & GGU & Gly G &   AGU & Ser S    \\
CGG & Arg R &  UGG & Trp W & GGG & Gly G &  AGG &  Arg R   \\
CGA & Arg R &  UGA & Stop & GGA & Gly G &   AGA & Arg R  \\[1mm] \hline
CAC & His H &   UAC & Tyr Y & GAC & Asp D &  AAC & Asn N  \\
CAU & His H &   UAU & Tyr Y &GAU & Asp D &   AAU & Asn N   \\
CAG & Gln Q &   UAG & Stop & GAG & Glu E &   AAG & Lys K   \\
CAA & Gln Q &   UAA & Stop & GAG & Glu E & AAG & Lys K   \\[1mm] \hline
\end{tabular}
\end{center}
\end{table}
  The pattern of codon usage varies between species
and even among tissues within a species.  Systematic studies for eukaryotes are rather fragmentary
 \citep{GG1982, ike1985, aota86, ikemura88, sharp1989, bulmer1991, huynen1992, akashi1994}.
The case of bacteriae has been widely studied \citep{bernardi85,bernardi86, osawa87,osawa88}.

 Some years ago a correlation between suitable ratios of codon usage
frequencies for  the synonimous quartets (sextets are considered as sum of a quartet and o doublet)
has been remarked in \citep{FSS1} for biological species
belonging to the vertebrate class and in \citep{CFSS} for biological
organisms including plants. 
It has  also  been observed that the previously defined ratios exhibits 
 an almost universal behaviour, i.e. independent from the
biological species and on the nature of the amino-acid.
Such correlations fit well in a mathematical
model of the genetic code, called {\it crystal basis model}, proposed  in \citep{FSS}. 

In this framework,  in \citep{FSS2}, it has been derived that the sum of the
codon usage probabilities for codons belonging to quartets, in the
generalized meaning specified above, with third nucleotide $C$ and $A$ (or
$G$ and $U$), should be a constant (sum rule) and in \citep{FSS3}  further
 investigation  on the statistical reliability of the observed pattern has been carried out.
 The observed pattern fits well in the {\it crystal basis model}, however we shall not discuss in detail the model and how the results fit in, as the main aim of this paper is to present the observed data, which may be an input for further study. We report in Appendix a brief summary of the model and the interested reader can find details in the quoted references.
 
  The aim of this paper is four-fold:
\begin{itemize}
\item to extend the previous analysis to species belonging to invertebrates and plants. For completeness and to make  relative comparisons, analysis for species belonging to vertebrates  is also reported, with an updated statistics;
\item to analyse specimens encoded by mitochondrial code;
\item  to check the reliability of  our analysis. This is done directly by improving the statistical tests  and indirectly by verifying a  consequence on the Shannon entropy derived theoretically by the sum rules.
\item to  search for correlation, between the  usage frequencies of codons, with the same final nucleotide,  encoding different amino-acid.   
\end{itemize}

The paper is organized as follows. In Sec. 2 we present the analysis of the codon usage frequencies (c.u.f.) for  eukaryotes belonging to the vertebrates, plants and invertebrates and for a specimen of mitochondria. In order to check the statistical reliability of our analysis for any specimen we perform the $\rho$  test  of control. In Sec.3 we   compare the theoretically  predicted behaviour of the Shannon entropy for  different biological species with the same compositional percentage of the same a.a. with the  values computed on the basis of the observed c.u.f.. In Sec. 4 we look for correlation between different amino acids. In Sec 5 we summarise and discuss the results.

\section{Analysis of the codon usage frequencies}

We define the usage probability or usage frequency for the codon $XZN$ ($N \in \{A,C, G, U\}$, 
$XZ \in \{CC, UC, GC, AC, CU, GU, CG. GG\}$) as

\begin{equation}
\label{eq:1}
P(XZN) = \lim_{n_{tot} \to \infty} \;\;\; \frac{n_{XZN}}{n_{XZ}}
\end{equation}
where $n_{XZN}$ is the number of times the codons $XZN$ has been used in
the biosynthesis process of the corresponding amino-acid and $n_{XZ}$ is
the total number of codons used to synthetise this amino-acid
\be
n_{tot} = \sum_{N=A,C, G, U} \, n_{XZN}
\ee
Note that we consider eight quartets, as we consider also the quartet sub-part of the three sextets, i.e. the  set of the four codons which differ only for the third nucleotide. In the following, to simplify the notation, we denote, for any fixed dinucleotide $XZ$, the probability  $P_N \equiv P(XZN)$, normalised 
\be
\sum_{ N = A,C, G, U} \; P_N = 1 \label{eq:n}
\ee
 As we compute the probabilities from observed frequencies, it follows that our analysis and predictions is restricted for biological species with enough
large statistics of codons.  In the following Subsections we compute the codon usage frequencies for the eight quartets,  for biological species belonging to the vertebrates, plants, invertebrates   and mitochondria.

 Our specimen is formed by   species with a codon statistics \citep{data} larger than 100,000 codons, for  vertebrates, plants and invertebrates, and larger than 30,000 codons  for mitochondria, see Tables
 \ref{tabledata1},  \ref{tabledata2},  \ref{tabledata3}, \ref{tabledata1m} (release 149.0 of  2005).

  For every specimen, 
 we define the average probability  ($n$ being the number of  biological species  in the specimen):
\be
<P_{N}>= \frac{1}{n} \sum_{i=1}^{n} P_{N,i} \quad N=
(A,C,G.U)
\label{eq:PN}
\ee
For completeness, we define in Appendix 6.2 the statistical quantities used in the following.

In order to test the hypotheses concerning the correlation coefficient $\rho$, we make use of the $t$ distribution \citep{Bryant}
\be
t = \frac{r \, \sqrt{n - 2}}{\sqrt{1 - r^{2}}}
\label{eq:t}
\ee
Like any statistical quantity computed from data the correlation coefficient $r$ differ from its "true value" $\rho$. The $t$ test can only be used to reject the hypothesis $\rho = 0$, i.e. no correlation. For any specimen, we compute the critical value $t_c$ such that:

- for $t(r) \in [t_c; -t_c]$ the probability that  $\rho = 0$ (uncorrelated variablel) is 95 \%

- for $t(r) >  t_c $ or $t(r) < -t_c $ the probability that  $\rho = 0$  is 5 \%.

In order to test the hypothesis  of the existence of a sum rule, we compute the standard deviation 
$\sigma(P_N+ P_{NÕ}) \equiv \sigma_{(N+N')}$ and we compare the computed value with the "theoretical" value 
 for two independent normally distributed variables $N$ and $N$, that is
\be
\sigma_{th}^2(N+N')=\sigma^2_{N} + \sigma^2_{N'} \label{eq:sth}
\ee
We  expect the value of the standard deviation of the sum of $P_N$ + $P(N')$ to be smaller than the sum of the two corresponding standard deviation,  as, due to the normalization condition eq.(\ref{eq:n}), the variables are not independent
\be
\sigma^2(P_{N+N'}) < \sigma^2(P_{N})+\sigma^2(P_{N'}) \;\;\;\;\;\;\;\;  (P_{N+N} \equiv P_{N}+P_{N'})
\ee
However we expect, in absence of any  further specific correlation, the reduction to be approximately equal for any couple  
$P_{N}$ and $P_{N'}$ extracted in the set of the four probabilities 
\begin{center}
\{$P_A \equiv P(XZA)$, $P_C \equiv P(XZC$), $P_G \equiv P(XZG)$, $P_U \equiv P(XZU)$\}
\end{center}
 while we discover  a reduction of the standard deviation for the sum 
$P_{A}+P_{C}$ and $P_{U}+P_{G}$ larger than for the other couples of probabilities. 

In Table \ref{eukca} and in Table   \ref{eukgu} we report the mean value and the standard deviation of the usage probabilities of the codon $XZN$ computed ovel all the biological species given in Tables \ref{tabledata1},  \ref{tabledata2} and \ref{tabledata3}. It can be remarked that the probability shows a rather large spread which is  reduced when one makes the sum. In Tables \ref{rid.var.ver}, \ref{rid.var.pln}, \ref{rid.var.inv} and \ref{rid.var.mit} we report, for any a.a., and for any specimen, the ratio of the measured $\sigma^2$ over the theoretical $\sigma^2_{th}$, defined in eq.(\ref{eq:sth}),  for the sum of the probabilities specified in the first column. In the last column the value of the ratio averaged over all a.a.is reported.

It can be remarked that this probability shows a rather
large spread which is surprisingly reduced when one makes the sum (compare
Tables \ref{rid.var.ver} and \ref{tablestat}).

 \begin{center} 
 \begin{table}[htbp]
 \footnotesize
 \begin{tabular}{|c||||c||c|c|c|c|c|c|c|c|}
 \hline
 b.sp. &  $<P_{N}>$ &  Arg & Leu & Ser & Thr & Pro & Ala & Gly & Val \\
\hline \hline
{\color{red}VRT} & $<P_{A}>$& {\color{red}  0,192} & {\color{red}  0,083} &  {\color{red}  0,223} &  {\color{red} 0,269 }&  {\color{red} 0,273 }& {\color{red}  0,221} & {\color{red}  0,269} &  {\color{red}  0,105} \\
& $\sigma_{P_{A}}$ & {\color{red}  0,024}  &{\color{red}  0,020}  & {\color{red}  0,026} &   {\color{red} 0,037} &  {\color{red} 0,036} & {\color{red}  0,037 } & {\color{red} 0,038}  & {\color{red} 0,026}\\
\hline
{\color{green}  PLN }& $<P_{A}>$   & {\color{green}    0,232}  &    {\color{green}    0,142}  &      {\color{green}  0,247}  &      {\color{green}  0,252}  &   {\color{green}     0,302}  &     {\color{green}   0,238}   &   {\color{green}    0,263}  &      {\color{green}  0,129}   \\
& $\sigma_{P_{A}}$    &    {\color{green}    0,114}  &     {\color{green}   0,066}  &   {\color{green}     0,088}   &     {\color{green}  0,074}  &   {\color{green}     0,116}  &    {\color{green}    0,082}    &    {\color{green}  0,089} &       {\color{green}  0,054}  \\
\hline
{\color{blue}INV }& $<P_{A}>$&  {\color{blue}  0,237}   &   {\color{blue} 0,154}   &  {\color{blue}  0,283}   &   {\color{blue} 0,305}  &   {\color{blue} 0,357}   & {\color{blue}   0,282}   & {\color{blue}    0,328 }  &  {\color{blue}  0,181}   \\
& $\sigma_{P_{A}}$&  {\color{blue}  0,113}   & {\color{blue}   0,082}   &  {\color{blue}  0,151}  &  {\color{blue}  0,111}   & {\color{blue}   0,167}   & {\color{blue}   0,115}   & {\color{blue}   0,141}   &  {\color{blue}  0,098}   \\
\hline \hline
{\color{red}VRT} & $<P_{C}>$& {\color{red}  0,329} & {\color{red}  0,250} &  {\color{red}  0,371} &  {\color{red} 0,371 }&  {\color{red} 0,319 }& {\color{red}  0,387} & {\color{red}  0,325} &  {\color{red}  0,253} \\
& $\sigma_{P_{C}}$ & {\color{red}  0,034}  &{\color{red}  0,023}  & {\color{red}  0,040} &   {\color{red} 0,049} &  {\color{red} 0,044} & {\color{red}  0,053 } & {\color{red} 0,038}  & {\color{red} 0,027}\\
\hline
{\color{green}  PLN }& $<P_{C}>$   & {\color{green}    0,272}  &    {\color{green}    0,278}  &      {\color{green}  0,261}  &      {\color{green}  0,297}  &   {\color{green}     0,223}  &     {\color{green}   0,274}   &   {\color{green}    0,277}  &      {\color{green}  0,264}   \\
& $\sigma_{P_{C}}$    &    {\color{green}    0,128}  &     {\color{green}   0,097}  &   {\color{green}     0,085}   &     {\color{green}  0,105}  &   {\color{green}     0,109}  &    {\color{green}    0,102}    &    {\color{green}  0,138} &       {\color{green}  0,101}  \\
\hline
{\color{blue}INV} & $<P_{C}>$&  {\color{blue}  0,273}   &   {\color{blue} 0,224}   &  {\color{blue}  0,240}   &   {\color{blue} 0,242}  &   {\color{blue} 0,211}   & {\color{blue}   0,261}   & {\color{blue}    0,267 }  &  {\color{blue}  0,220}   \\
& $\sigma_{P_{C}}$&  {\color{blue}  0,158}   & {\color{blue}   0,078}   &  {\color{blue}  0,105}  &  {\color{blue}  0,106}   & {\color{blue}   0,116}   & {\color{blue}   0,130}   & {\color{blue}   0,176}   &  {\color{blue}  0,095}   \\
\hline \hline
{\color{red}VRT} & $<P_{C+A}>$& {\color{red}  0,521} & {\color{red}  0,333} &  {\color{red}  0,594} &  {\color{red} 0,641 }&  {\color{red} 0,592 }& {\color{red}  0,609} & {\color{red}  0,594} &  {\color{red}  0,358} \\
& $\sigma_{P_{C+A}}$ & {\color{red}  0,016}  &{\color{red}  0,013}  & {\color{red}  0,019} &   {\color{red} 0,021} &  {\color{red} 0,020} & {\color{red}  0,027 } & {\color{red} 0,013}  & {\color{red} 0,015}\\
\hline
{\color{green}  PLN }& $<P_{C+A}>$   & {\color{green}    0,510}  &    {\color{green}    0,378}  &      {\color{green}  0,523}  &      {\color{green}  0,547}  &   {\color{green}     0,567}  &     {\color{green}   0,543}   &   {\color{green}    0,595}  &      {\color{green}  0,401}   \\
& $\sigma_{P_{C+A}}$    &    {\color{green}    0,135}  &     {\color{green}   0,081}  &   {\color{green}     0,071}   &     {\color{green}  0,060}  &   {\color{green}     0,106}  &    {\color{green}    0,076}    &    {\color{green}  0,121} &       {\color{green}  0,064}  \\
\hline
{\color{blue}INV} & $<P_{C+A}>$&  {\color{blue}  0,504}   &   {\color{blue} 0,420}   &  {\color{blue}  0,508}   &   {\color{blue} 0,549}  &   {\color{blue} 0,525}   & {\color{blue}   0,512}   & {\color{blue}    0,540 }  &  {\color{blue}  0,393}   \\
& $\sigma_{P_{C+A}}$&  {\color{blue}  0,086}   & {\color{blue}   0,067}   &  {\color{blue}  0,044}  &  {\color{blue}  0,043}   & {\color{blue}   0,048}   & {\color{blue}   0,037}   & {\color{blue}   0,081}   &  {\color{blue}  0,077}   \\
\hline
  \end{tabular} \centering\caption{\footnotesize{Average value of $P_C$, $P_A$ and of their sum with the corresponding standard deviation for the 8 amino-acids encoded by quartets.}}\label{eukca}
 \bigskip
 \begin{tabular}{|c||||c||c|c|c|c|c|c|c|c|}
 \hline
 b.sp. &  $<P_{N}>$ &  Arg & Leu & Ser & Thr & Pro & Ala & Gly & Val \\
\hline \hline
{\color{red}VRT} & $<P_{U}>$& {\color{red}  0,170} & {\color{red}  0,155} &  {\color{red}  0,309} &  {\color{red} 0,237 }&  {\color{red} 0,287 }& {\color{red}  0,280} & {\color{red}  0,178} &  {\color{red}  0,176} \\
& $\sigma_{P_{U}}$ & {\color{red}  0,045}  &{\color{red}  0,035}  & {\color{red}  0,029} &   {\color{red} 0,028} &  {\color{red} 0,024} & {\color{red}  0,031 } & {\color{red} 0,026}  & {\color{red} 0,037}\\
\hline
{\color{green}  PLN }& $<P_{U}>$   & {\color{green}    0,304}  &    {\color{green}    0,313}  &      {\color{green}  0,305}  &      {\color{green}  0,286}  &   {\color{green}     0,304}  &     {\color{green}   0,331}   &   {\color{green}    0,310}  &      {\color{green}  0,320}   \\
& $\sigma_{P_{U}}$    &    {\color{green}    0,114}  &     {\color{green}   0,125}  &   {\color{green}     0,075}   &     {\color{green}  0,087}  &   {\color{green}     0,074}  &    {\color{green}    0,095}    &    {\color{green}  0,092} &       {\color{green}  0,108}  \\
\hline
{\color{blue}INV }& $<P_{U}>$&  {\color{blue}  0,343}   &   {\color{blue} 0,301}   &  {\color{blue}  0,261}   &   {\color{blue} 0,257}  &   {\color{blue} 0,234}   & {\color{blue}   0,293}   & {\color{blue}    0,288 }  &  {\color{blue}  0,305}   \\
& $\sigma_{P_{U}}$&  {\color{blue}  0,191}   & {\color{blue}   0,182}   &  {\color{blue}  0,098}  &  {\color{blue}  0,113}   & {\color{blue}   0,110}   & {\color{blue}   0,123}   & {\color{blue}   0,137}   &  {\color{blue}  0,139}   \\
\hline \hline
{\color{red}VRT} & $<P_{G}>$& {\color{red}  0,308} & {\color{red}  0,512} &  {\color{red}  0,096} &  {\color{red} 0,123 }&  {\color{red} 0,121 }& {\color{red}  0,111} & {\color{red}  0,229} &  {\color{red}  0,466} \\
& $\sigma_{P_{G}}$ & {\color{red}    0,045}  &{\color{red}  0,035}  & {\color{red}  0,017} &   {\color{red} 0,026} &  {\color{red} 0,027} & {\color{red}  0,023 } & {\color{red} 0,030}  & {\color{red} 0,040}\\
\hline
{\color{green}  PLN }& $<P_{G}>$   & {\color{green}    0,193}  &    {\color{green}    0,267}  &      {\color{green}  0,187}  &      {\color{green}  0,165}  &   {\color{green}     0,171}  &     {\color{green}   0,157}   &   {\color{green}    0,150}  &      {\color{green}  0,287}   \\
& $\sigma_{P_{G}}$    &    {\color{green}    0,128}  &     {\color{green}   0,097}  &   {\color{green}     0,085}   &     {\color{green}  0,105}  &   {\color{green}     0,109}  &    {\color{green}    0,102}    &    {\color{green}  0,138} &       {\color{green}  0,101}  \\
\hline
{\color{blue}INV} & $<P_{G}>$&  {\color{blue}  0,147}   &   {\color{blue} 0,321}   &  {\color{blue}  0,216}   &   {\color{blue} 0,196}  &   {\color{blue} 0,198}   & {\color{blue}   0,164}   & {\color{blue}    0,117 }  &  {\color{blue}  0,294}   \\
& $\sigma_{P_{G}}$&  {\color{blue}  0,091}   & {\color{blue}   0,203}   &  {\color{blue}  0,128}  &  {\color{blue}  0,119}   & {\color{blue}   0,133}   & {\color{blue}   0,106}   & {\color{blue}   0,071}   &  {\color{blue}  0,160}   \\
\hline \hline
{\color{red}VRT} & $<P_{U+G}>$              & {\color{red} 0,479} & {\color{red} 0,667}  & {\color{red} 0,406} & {\color{red} 0,359} &  {\color{red}0,408} &  {\color{red} 0,391} & {\color{red} 0,406} & {\color{red} 0,642}\\
&$\sigma(P_{U+G})$          & {\color{red} 0,016} & {\color{red} 0,013}  & {\color{red} 0,019} & {\color{red} 0,021} & {\color{red}  0,020}&  {\color{red} 0,027} & {\color{red} 0,013} & {\color{red} 0,015} \\ \hline
{\color{green}  PLN} & $<P_{U+G}>$     &   {\color{green} 0,49}    &   {\color{green} 0,622}   &    {\color{green} 0,477}   &  {\color{green}  0,453}   &  {\color{green}  0,433}   &   {\color{green} 0,457}   &    {\color{green} 0,405}   &   {\color{green} 0,599}   \\
& $\sigma(P_{U+G})$           &   {\color{green}  0,135}   &  {\color{green}   0,081}   &   {\color{green}  0,071}   &  {\color{green}  0,060}    &   {\color{green}  0,106}   &   {\color{green}  0,076}   &   {\color{green}  0,121}   &  {\color{green}  0,064}   \\
\hline 
{\color{blue}INV} &  $<P_{U+G}>$     &  {\color{blue}  0,496}   & {\color{blue}   0,580}    &  {\color{blue}  0,492}   &  {\color{blue}  0,451}   &  {\color{blue}  0,475}   &  {\color{blue}  0,488}   & {\color{blue}   0,460}    &  {\color{blue}  0,607}   \\
& $\sigma(P_{U+G})$      & {\color{blue}   0,086}   & {\color{blue}   0,067}   &  {\color{blue}  0,044}   &   {\color{blue} 0,043}   & {\color{blue}   0,048}   &  {\color{blue}  0,037}   & {\color{blue}  0,081}   & {\color{blue}   0,077}   \\
\hline
  \end{tabular}\centering\caption{\footnotesize{Average value of $P_G$, $P_U$ and of their sum with the corresponding standard deviation for the 8 amino-acids encoded by quartets.}} \label{eukgu}
   \end{table}
 \end{center}
\begin{table}[htbp]
\begin{tabular}{|c||c|c|c|c|c|c|c|c||c|} \hline
\texttt{VRT}.&R&L&S&T&P&A&G&V&$\small{<{\sigma_{mis}}^2/
{\sigma_{th}}^2>_{a.a}}$\\
\hline
$ P_{C+A}$    &   0,15    &   0,18    &   0,16    &   0,11    &   0,13    &   0,18    &   0,06    &   0,16    &   0,14    \\
$P_{U+G}$    &   0,07    &   0,07    &   0,32    &   0,29    &   0,32    &   0,50    &   0,12    &   0,08    &   0,22    \\
$P_{U+C}$    &   0,63    &   0,20    &   0,17    &   0,23    &   0,37    &   0,23    &   0,30    &   0,35    &   0,31    \\
$P_{A+G}$    &   0,77    &   0,22    &   0,44    &   0,36    &   0,47    &   0,46    &   0,28    &   0,32    &   0,42    \\
$P_{U+A}$    &   1,35    &   1,75    &   1,74    &   1,87    &   1,71    &   1,75    &   1,72    &   1,64    &   1,69    \\
$P_{C+G}$    &   1,11    &   1,64    &   1,37    &   1,33    &   1,19    &   1,22    &   1,54    &   1,49    &   1,36    \\
\hline
\end{tabular}
\centering \caption {\footnotesize{Value of the ratio
$\sigma^2/{\sigma^2}_{th}$ for the sum of probabilities computed over the specimen of 
\textbf{vertebrates}; in the last column $<\sigma^{2}/{\sigma_{th}}^2>_{a.a}$ is the average value computed over the  8
aminoacids.}}\label{rid.var.ver}
\bigskip
\begin{tabular}{|c||c|c|c|c|c|c|c|c||c|} \hline
\texttt{PLN.}&R&L&S&T&P&A&G&V&$\small{<{\sigma_{mis}}^2/
{\sigma_{th}}^2>_{a.a}}$\\
\hline
$P_{C+A}$   &   0,25    &   0,33    &   0,13    &   0,12    &   0,09    &   0,08    &   0,25    &   0,45    &   0,21    \\
$P_{U+G}$   &   0,41    &   0,15    &   0,15    &   0,15    &   0,20    &   0,09    &   0,61    &   0,28    &   0,26    \\
$P_{U+C}$   &   0,34    &   0,45    &   0,26    &   0,23    &   0,49    &   0,28    &   0,32    &   0,26    &   0,33    \\
$P_{A+G}$   &   0,55    &   0,57    &   0,22    &   0,42    &   0,44    &   0,43    &   0,88    &   0,45    &   0,50    \\
$P_{U+A}$   &   1,41    &   1,17    &   1,56    &   1,44    &   1,46    &   1,53    &   1,04    &   1,15    &   1,35    \\
$P_{C+G}$   &   1,16    &   1,45    &   1,70    &   1,74    &   1,37    &   1,62    &   1,35    &   1,57    &   1,50    \\
\hline
\end{tabular}
\centering \caption{\footnotesize{Value of the ratio
$\sigma^2/{\sigma^2}_{th}$ for the sum of probabilities computed over the specimen of 
\textbf{plants}; in the last column $<\sigma^{2}/{\sigma_{th}}^2>_{a.a}$ is the average value computed over the  8
aminoacids.}} \label{rid.var.pln}
\bigskip
\begin{tabular}{|c||c|c|c|c|c|c|c|c||c|} \hline
\texttt{INVRT}.&R&L&S&T&P&A&G&V&$\small{<{\sigma_{mis}}^2/
{\sigma_{th}}^2>_{a.a}}$\\
\hline
$P_{C+A}$   &   0,48    &   0,51    &   0,15    &   0,15    &   0,27    &   0,19    &   0,29    &   0,22    &   0,28    \\
$P_{U+G}$   &   0,41    &   0,09    &   0,20    &   0,14    &   0,38    &   0,22    &   0,61    &   0,09    &   0,27    \\
$P_{U+C}$   &   0,25    &   0,66    &   0,29    &   0,26    &   0,50    &   0,25    &   0,44    &   0,38    &   0,38    \\
$P_{A+G}$   &   0,74    &   0,54    &   0,16    &   0,24    &   0,28    &   0,33    &   0,89    &   0,31    &   0,44    \\
$P_{U+A}$   &   0,99    &   1,37    &   1,50    &   1,62    &   1,18    &   1,51    &   0,99    &   1,68    &   1,36    \\
$P_{C+G}$   &   1,46    &   1,16    &   1,78    &   1,60    &   1,51    &   1,52    &   1,05    &   1,41    &   1,44    \\
\hline
\end{tabular}
\centering \caption{\footnotesize{Value of the ratio
$\sigma^2/{\sigma^2}_{th}$ for the sum of probabilities computed over the specimen of 
\textbf{invertebrates}; in the last column $<\sigma^{2}/{\sigma_{th}/}^2>_{a.a}$ is the average value computed over the  8
aminoacids.}}\label{rid.var.inv}
\bigskip
\begin{tabular}{|c||c|c|c|c|c|c|c|c||c|}
\hline \texttt{Mit.}&R&L&S&T&P&A&G&V&$\small{<{\sigma_{mis}}^2/
{\sigma_{th}}^2>_{a.a}}$\\
\hline
$P_{C+A}$&   0,30    &   0,66    &   0,17    &   0,16    &   0,23    &   0,27    &   0,31    &   0,42    &   0,32    \\
$P_{A+G}$&   0,73    &   0,38    &   0,55    &   0,80    &   0,72    &   0,62    &   0,67    &   0,47    &   0,62    \\
$P_{C+U}$&   0,84    &   0,36    &   0,57    &   0,74    &   0,87    &   0,39    &   0,67    &   0,56    &   0,63    \\
$P_{U+A}$&   0,73    &   0,53    &   0,95    &   0,81    &   0,58    &   1,10    &   0,75    &   0,60    &   0,76    \\
$P_{C+G}$&   0,83    &   1,09    &   1,27    &   0,93    &   0,95    &   0,93    &   0,81    &   1,18    &   1,00     \\
$P_{U+G}$&   0,88    &   0,92    &   0,90    &   1,16    &   1,15    &   0,92    &   1,54    &   0,94    &   1,05    \\
\hline
\end{tabular}
\centering \caption{\footnotesize{Value of the ratio
$\sigma^2/{\sigma^2}_{th}$ for the sum of probabilities computed over the specimen of 
\textbf{mitochondria}; in the last column $<\sigma^{2}/{\sigma_{th}}^2>_{a.a}$ is the average value computed over the  8
aminoacids.}} \label{rid.var.mit}
\end{table}.

\newpage

\subsection{Correlation evaluation}

We  evaluate the symmetric correlation matrix defined by eq.(\ref{eq:rXY}) ($\small{XZ= CC,UC,CG,AC,CU,GU,GC,GG}$)

\begin{center}
\begin{tabular}{|c|c|c|c|c|}
 \hline
 & XZA &XZC &XZG &XZU \\
 \hline
 XZA & 1 & $r_{CA}$ & $r_{AG}$ & $r_{UA}$ \\
 \hline
 XZC &$r_{CA}$ & 1 & $r_{CG}$ & $r_{UC}$ \\
 \hline
XZG & $r_{AG}$ & $r_{CG}$ & 1 & $r_{UG}$  \\
\hline
 XZU & $r_{UA}$&$r_{UC}$ & $r_{UG}$ & 1 \\
 \hline
\end{tabular}
\end{center}
 where, e.g. if  $XZ=CC$
 $$r_{UC} \equiv \left(P(CCU),P
(CCC)\right) \equiv \left((P_1(CCU)....P_n(CCU));(P_1(CCC)....P_n(CCC))\right)$$
and
 $P_{i}(CCN)$ is the observed frequency for  the codon  $CCN$ in the $i-th$ species of the specimen. 
 Considering  the 8 quartets we compute a set of   48 numerical coefficients from which we can
 extract information about a possible pattern of regularity.
 It may be useful to recall the following table 
\begin{center}
\begin{tabular}{c||l} \hline
 Absolute value  of $r$ &  Correlation degree \\
\hline
    0,00-0,20  &    Very low\\
    0,20-0,40  &    Low \\
    0,40-0,60  &   Average \\
    0,60-0,80  &   High \\
    0,80-1,00  &   Very high \\
\hline
\end{tabular}
\end{center}

\subsection{Vertebrates}

In Table \ref{r-ver} we report the  value of the correlation matrix for the six independent couples of probabilities. 
\begin{table}[htbp]
\begin{tabular}{|c|r|r|r|r|r|r|r|r|r|}
\hline
$r_{XY}$&$r_{CA}$&$r_{UG}$&&$r_{UC}$&$r_{AG}$&&$r_{UA}$&$r_{CG}$\\
\hline \hline
P   &   -0,89   &   -0,69   & &  -0,75   &   -0,55   &&   0,76    &   0,21    \\
T   &   -0,92   &   -0,71   & &  -0,89   &   -0,68   &&   0,91    &   0,40   \\
A   &   -0,88   &   -0,53   & &  -0,89   &   -0,60   &&   0,76    &   0,30   \\
S   &   -0,92   &   -0,77   & &  -0,87   &   -0,60   &&   0,75    &   0,51    \\
V   &   -0,84   &   -0,93   & &  -0,69   &   -0,74   &&   0,68    &   0,53    \\
L   &   -0,83   &   -0,93   & &  -0,87   &   -0,91   &&   0,87    &   0,69    \\
R   &   -0,90   &   -0,93   & &  -0,39   &   -0,27   &&   0,41    &   0,11    \\
G   &   -0,94   &   -0,89   & &  -0,75   &   -0,74   &&   0,77    &   0,56    \\
\hline
$<r>_{a.a}$ &   -0,89   &   -0,80    &&   -0,76   &   -0,64   & &  0,74    &   0,41    \\
\hline \hline
\end{tabular}
\centering \caption{Specimen of \textbf{vertebrates}:
value of the correlation coefficient for the 6 independent couples of probabalities for the  8
quartets.  In the last row the value of the coefficient averaged over the 8 a.a..} \label{r-ver}
\end{table}

 From Table \ref{r-ver} we remark a clear anti-correlation, for all a.a., between $P_A$ and $P_C$. A comparable degree of anti-correlation appears for the complementary couple $P_U - P_G$ for the a.a. Val, Leu, Arg and Gly. Let us also remark that an anti-correlation  in the couples $P_U-P_C$ and $P_A - P_G$, naively expected as the concerned nucleotides belong to the same family (respectively, pyrimidine and purine), is present, but not in all the a.a..

\begin{figure}[htbp]
\centering 
\includegraphics
[width=16cm]{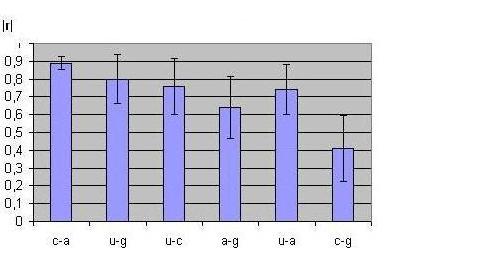} \\
 \caption{Specimen of \textbf{vertebrates}: Absolute value of the six highest correlation coefficient  averaged over 8 a.a.. The error is computed by a  standard deviation.}\label{vertcorr.jpg}
%
\includegraphics[width=16cm]{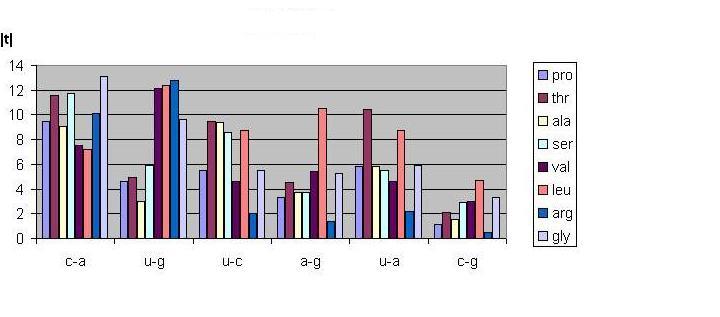}
\centering\caption{Value of  $|t|$ for the correlation coefficients for the specimen of vertebrates}
 \label{figure/TOVRT.jpg}
 \end{figure}
 
 \subsubsection[]{TEST  $\rho=0$  for VERTEBRATES (n=26)}

From eq.(\ref{eq:t}), we get  the critical value of the statistical label $t$ at the 95\%  confidence level 
 
  $${t_{c}=t_{5\%}^{n-2}=t_{0,025}^{24}=2,06}$$

\begin{itemize}
    \item    for $t(r) \in [2,06;  -2,06]$ the probability that $\rho=0$ (uncorrelated variables) 
    is 95\%
 \item  for $t(r)> t_c =2,06 \quad \mbox{or} \quad t_{r}<-t_c =-2,06$ 
        the probability that che $\rho=0$ is 5\%.
\end{itemize}
In Tables \ref{t-ver} and \ref{Int.vrt} we report, respectively,  the computed value of the $t$ label for any a.a. and the interval where the values of the correlation coefficient $r$ are included in, at 95 \% confidence level.
\begin{table}[h]
\begin{tabular}{|c|l|l|l|l|l|l|l|}
\hline
$VRT$&$r_{CA}$&$r_{UG}$&$r_{UC}$&$r_{AG}$&$r_{UA}$&$r_{CG}$\\
\hline \hline
t[P]  &   -9,5    &   -4,6    &   -5,5    &   -3,3    &   5,8            &   \textbf{1,1} \\
t[T]   &   -11,6   &   -4,9    &   -9,5    &   -4,5    &   10,4           &   2,1 \\
t[A]   &   -9,1    &   -3,0    &   -9,4    &   -3,7    &   5,8            &   \textbf{1,5} \\
t[S]   &   -11,7   &   -5,9    &   -8,6    &   -3,7    &   5,5            &   2,9 \\
t[V]   &   -7,5    &   -12,1   &   -4,6    &   -5,4    &   4,6            &   3,0   \\
t[L]   &   -7,2    &   -12,4   &   -8,7    &   -10,5   &   8,7            &   4,7 \\
t[R]   &   -10,1   &   -12,8   &   \textbf{-2,0}       &  \textbf{-1,4}   &   \textbf{2,2} &   \textbf{0,5} \\
t[G]   &   -13,1   &   -9,6    &   -5,5                &   -5,3           &   5,9 &   3,3 \\
\hline
$<t>_{a.a.}$&   -9,98   &   -8,16   &   -6,73   &   -4,73   &   6,11    &   2,39    \\
\hline \hline
\end{tabular}
\centering\caption{Specimen of \textbf{vertebrates}:
value of the statistical label $t$  for the  6  correlation coefficients for the 
 8 a.a.. In the last row the average value of   $t$.} \label{t-ver}

\begin{tabular}{|c|c|c|c|c|c|c|c|c|c|}
\hline
VRT &$r_{CA}$&$r_{UG}$&&     $r_{UC}$ & $r_{AG}$&      &$r_{UA}$&$r_{CG}$\\
\hline \hline
P   &    -0,77  -0,95  &  -0,41   -0,85   &&   -0,51   -0,88   &   -0,21   -0,77   &&   0,53    0,89    &   -0,19   0,55    \\
T   &    -0,83  -0,96  &  -0,45   -0,86   &&   -0,77   -0,95   &   -0,4    -0,84   &&   0,81    0,96    &   0,02    0,68    \\
A   &    -0,75  -0,95  &  -0,18   -0,76   &&   -0,77   -0,95   &   -0,28   -0,80   &&   0,53    0,89    &   -0,1    0,62    \\
S   &    -0,83  -0,96  &  -0,55   -0,89   &&   -0,73   -0,94   &   -0,28   -0,80   &&   0,51    0,88    &   0,15    0,75    \\
V   &    -0,67  -0,93  &  -0,85   -0,97   &&   -0,41   -0,85   &   -0,49   -0,88   &&   0,40    0,84    &   0,18    0,76    \\
L   &    -0,65  -0,92  &  -0,85   -0,97   &&   -0,73   -0,94   &   -0,81   -0,96   &&   0,73    0,94    &   0,41    0,85    \\
R   &    -0,79  -0,95  &  -0,85   -0,97   &&    0,00   -0,68   &   0,13    -0,60   &&   0,03    0,69    &   -0,29   0,48    \\
G   &    -0,87  -0,97  &  -0,77   -0,95   &&   -0,51   -0,88   &   -0,49   -0,88   &&   0,55    0,89    &   0,22    0,78    \\
\hline
\end{tabular}
\centering \caption{Range of the values of the correlation coefficient $r$,  at 95\%  confidence level, for the specimen of 26 vertebrates}
 \label{Int.vrt} 
\end{table}

From  Table \ref{t-ver} we remark that the values of the modulus of the statistical label $t$   computed over the specimen of vertebrates is, in the average, more far, as expected,  from the critical value
   $t_c=2,06$ for the correlation $r_{CA}$ and $r_{UG}$ than for $r_{UC}$ and $r_{AG}$. From  Table \ref{Int.vrt}, one correlation factor can be vanishing (95\%  confidence level)   for the couples   (U,C), (A,G),  i.e. Arg, and, at least, three for the couple  (C,G), i.e. Arg,  Pro, and Ala.

\subsection{Plants}

In Table \ref{r-pl} we report the  value of the correlation matrix for the six independent couples of probabilities. 
In Table \ref{rid.var.pln} we report the mean value and the standard
deviation of the usage probability   of the codons $XZN$ $(XZ = NC, CU,
GU, CG, GG)$ computed over all biological species given in Table
\ref{tabledata3}. It can be remarked that this probability shows a rather
large spread which is surprisingly reduced when one makes the sum (compare
Tables \ref{rid.var.pln} and \ref{tablestat}).

\begin{table}[htbp]
\begin{tabular}{|c|r|r|r|r|r|r|r|r|r|}
\hline
$r_{XY}$&$r_{CA}$&$r_{UG}$&& $r_{UC}$&$r_{AG}$&& $r_{UA}$&$r_{CG}$\\
\hline \hline
P   &   -0,91   &   -0,81   &&   -0,54   &   -0,61   &&   0,41    &   0,48    \\
T   &   -0,94   &   -0,87   &&   -0,79   &   -0,59   &&   0,75    &   0,48    \\
A   &   -0,94   &   -0,93   &&   -0,72   &   -0,57   &&   0,63    &   0,55    \\
S   &   -0,87   &   -0,86   &&   -0,75   &   -0,78   &&   0,71    &   0,56    \\
V   &   -0,66   &   -0,72   &&   -0,75   &   -0,65   &&   0,71    &   0,15    \\
L   &   -0,72   &   -0,85   &&   -0,57   &   -0,52   &&   0,54    &   0,17    \\
R   &   -0,76   &   -0,66   &&   -0,67   &   -0,50   &&   0,16    &   0,49    \\
G   &   -0,83   &   -0,48   &&   -0,73   &   -0,14   &&   0,36    &   0,07    \\
\hline
$<r>_{a.a.}$ &   -0,83   &   -0,77   &&   -0,69   &   -0,55   &&   0,53    &   0,37    \\
\hline \hline
\end{tabular}
 \centering \caption {Specimen of \textbf{plants}:
value of the correlation coefficient for the 6 independent couples of probabilities for the  8
quartets.  In the last row the value of the coefficient averaged over the 8 a.a..} \label{r-pl}
\end{table}

From Table \ref{r-pl} we remark a clear anti-correlation, for all a.a., between $P_A$ and $P_C$ and between $P_U$ and $P_G$ for the a.a. Pro, Thr, Ala, and Ser, while  $P_U -P_C$ and $P_A - P_G$ are weakly anti-correlated.

We remark the foliowing {\bf correlation pattern}:
\begin{itemize}
    \item P$\rightarrow |r_{CA}|>     |r_{UG}|>     |r_{AG}|>     |r_{UC}|>     |r_{CG}|>  |r_{UA}|$
    \item S$\rightarrow |r_{CA}|\simeq|r_{UG}|>     |r_{AG}|>     |r_{UC}|>     |r_{UA}|>  |r_{CG}|$

    \item A$\rightarrow |r_{CA}|\simeq|r_{UG}|>     |r_{UC}|>     |r_{UA}|>     |r_{AG}|>  |r_{CG}|$
    \item T$\rightarrow |r_{CA}|>     |r_{UG}|>     |r_{UC}|>     |r_{UA}|>     |r_{AG}|>  |r_{CG}|$

    \item R$\rightarrow |r_{CA}|>     |r_{UC}|\simeq|r_{UG}|>     |r_{CG}|\simeq|r_{AG}|>  |r_{UA}|$
    \item G$\rightarrow |r_{CA}|>     |r_{UC}|\simeq|r_{UG}|>     |r_{UA}|>     |r_{AG}|>  |r_{CG}|$

    \item V$\rightarrow |r_{UC}|>     |r_{UG}|\simeq|r_{UA}|>     |r_{CA}|\simeq|r_{AG}|>  |r_{CG}|$
    \item L$\rightarrow |r_{UG}|>     |r_{CA}|>     |r_{UC}|>     |r_{UA}|>     |r_{AG}|>  |r_{CG}|$
\end{itemize}
that is \emph{for almost all the quartets} the highest value of the correlation coefficient is {\bf always} between $P_{C}$ and $P_{A}$ or $P_{U}$ and $P_{G}$
(P, T, A, S, G, R and, respectively, L)  except for V, for which the highest value is $r_{UC}$, close to the value of $r_{UG}$ and $r_{CA}$. The lowest value for the correlation coefficient  is  between $P_{C}$  and $P_{G}$ or between $P_{U}$ and $P_{A}$ (P T A S R G,  resp. V and L).

Averaging over the 8 a.a. one gets: \\
$$<r_{CA}>_{a.a}= -0,83 \qquad <r_{UC}>_{a.a}= -0,69 \qquad <r_{UA}>_{a.a}= 0,53$$

$$<r_{UG}>_{a.a}= -0,77 \qquad <r_{AG}>_{a.a}= -0,55 \qquad <r_{CG}>_{a.a}=0,37$$

\begin{figure}[htbp]\centering
\includegraphics[width=16cm]{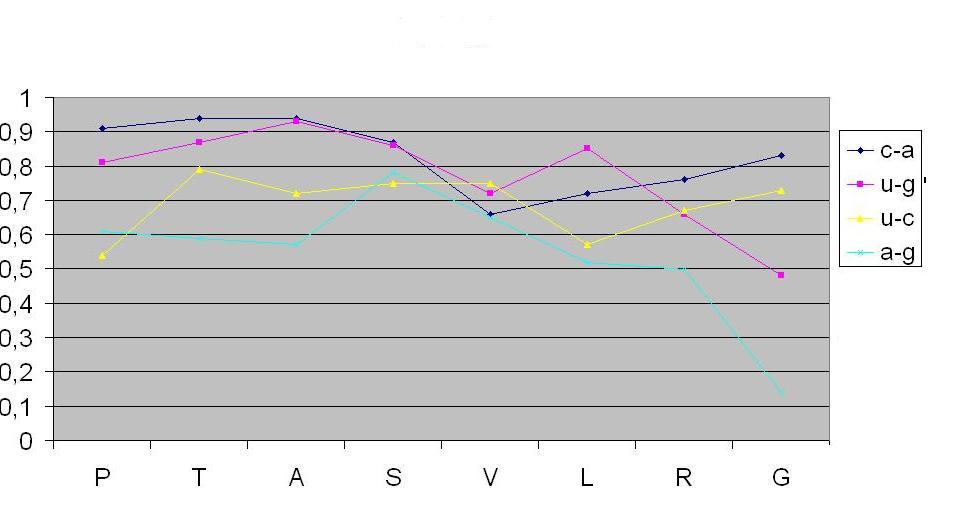}\\
\caption{Specimen of \textbf{plants}: absolute value of the six highest correlation coefficient  averaged over 8 a.a.. correlation coefficient between codon frequency usage for quartets ending in (C,A),(U,G),  (U,C) and (A,G). We do not report the correlations 
 for quartets ending in (U,A) and (C,G), which are in general very low.}
\label{piante.jpg}\
\end{figure}

\begin{figure}[htbp]\centering
\includegraphics[width=16cm]{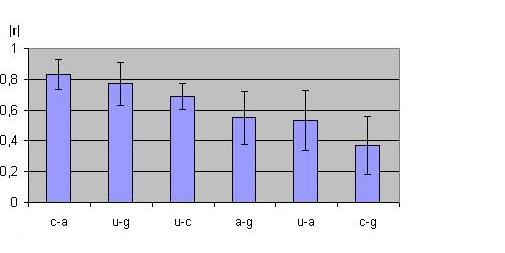}
  \caption{Specimen of \textbf{plants}: Absolute value of the six highest correlation coefficients  averaged over 8 a.a.. The error is computed by a  standard deviation.}\label{figure/plncorr.jpg}
\end{figure}

\subsubsection{Test $\rho=0$ for  plants (n = 38)}

 From eq.(\ref{eq:t}), we get that the critical value of the statistical label  $t$ (95\% confidence level) is
$${t_{c}=t_{5\%}^{n-2}=t_{0,025}^{36}=2,02}$$

\begin{itemize}
    \item   for $t_{r} \in [2,02; -2,02]$ the probability that $\rho=0$ (uncorrelated variables) 
       is 95\%.

    \item   for $t_{r}> 2,02 \quad$ or  $\quad
    t_{r}<-2,02$  the probability that $\rho=0$ is 5\%  
\end{itemize}
\begin{table}[htbp]
\begin{tabular}{|c|r|r|r|r|r|r|r|r|r|}
\hline
$PLN$  &$r_{CA}$&  $r_{UG}$ &&   $r_{UC}$&$r_{AG}$   &&  $r_{UA}$&$r_{CG}$\\
\hline \hline
t[P]   &   -13,4   &   -8,2    &&   -3,9    &   -4,6    &&   2,7 &   3,3 \\
t[T]   &   -16,6   &   -10,6   &&  -7,6     &   -4,3    &&   6,7 &   3,3 \\
t[A]  &   -16,8   &   -14,6   &&   -6,2    &   -4,2    &&   4,9 &   3,9 \\
t[S]  &   -10,7   &   -10,1   &&   -6,8    &   -7,6    &&   6,1 &   4,1 \\
t[V]   &   -5,3    &   -6,3    &&   -6,7    &   -5,1    &&   6,1 &   \textbf{0,9} \\
t[L]  &   -6,2    &   -9,8    &&   -4,1    &   -3,6    &&   3,9 &   \textbf{1,1} \\
t[R]   &   -6,9    &   -5,3    &&   -5,4    &   -3,5    &&   \textbf{1,0}   &   3,3 \\
t[G]  &   -8,9    &   -3,3    &&   -6,5    &   \textbf{-0,9}    &&   2,3 &   \textbf{0,4} \\
\hline
$<t>_{a.a.}$    &   10,6    &   8,5 &&   5,9 &   4,2 &&   4,2 &   2,5 \\
\hline \hline
\end{tabular}
\centering\caption{Specimen of \textbf{plants}:
value of the statistical label $t$  for the  6  correlation coefficients for the 
 8 a.a.. In the last row the average value of   $t$.} \label{t-pl}
\begin{tabular}{|c|c|c|c|c|c|c|c|c|c|}
\hline
PLN. &$r_{CA}$&$r_{UG}$&&     $r_{UC}$ & $r_{AG}$&&      $r_{UA}$&$r_{CG}$\\
\hline \hline
P   &   -0,83   -0,95   &   -0,66   -0,90   &&   -0,27   -0,73   &   -0,36   -0,78   &&   0,1     0,65    &   0,19    0,69    \\
T   &   -0,89   -0,97   &   -0,76   -0,93   &&   -0,63   -0,89   &   -0,33   -0,77   &&   0,57    0,86    &   0,19    0,69    \\
A   &   -0,89   -0,97   &   -0,87   -0,96   &&   -0,52   -0,85   &   -0,31   -0,75   &&   0,39    0,79    &   0,28    0,74    \\
S   &   -0,76   -0,93   &   -0,75   -0,93   &&   -0,57   -0,86   &   -0,61   -0,88   &&   0,50    0,84    &   0,29    0,75    \\
V   &   -0,43   -0,81   &   -0,52   -0,85   &&   -0,57   -0,86   &   -0,42   -0,8    &&   0,50    0,84    &   -0,18   0,45    \\
L   &   -0,52   -0,85   &   -0,73   -0,92   &&   -0,31   -0,75   &   -0,24   -0,72   &&   0,27    0,73    &   -0,16   0,46    \\
R   &   -0,58   -0,87   &   -0,43   -0,81   &&   -0,45   -0,82   &   -0,21   -0,71   &&   -0,17   0,46    &   0,2 0,7 \\
G   &   -0,69   -0,91   &   -0,19   -0,69   &&   -0,54   -0,85   &    0,19   -0,44   &&   0,05    0,61    &   -0,26   0,38    \\
\hline
\end{tabular}
\centering \caption{Confidence intervals   (95\%  confidence level) for the specimen of n=38 plants}\label{Int.pln}
\end{table}
  \begin{figure}[htbp]\centering
\includegraphics[width=16cm]{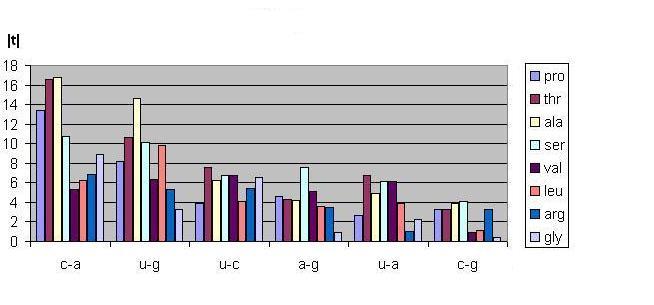}\\
\caption{Value of $|t|$ for the correlation coefficients for the specimen of plants.}\label{figure/T0pln.jpg}\end{figure}
In Tables \ref{t-pl} and \ref{Int.pln} we report, respectively,  the computed value of the $t$ label for any a.a. and the interval where the values of the correlation coefficient $r$ are included in, at 95 \% confidence level.

From  Table \ref{t-pl} we remark that the values of the modulus of the statistical label $t$    computed over the specimen of plants is, in the average, more far from the critical value
   $t_c=2,02$ for the correlation $r_{CA}$ and $r_{UG}$ than for $r_{UC}$ and $r_{AG}$. From Table \ref{Int.pln}, one correlation factor can be  vanishing (95\%  confidence level) for the couple  (A,G), i.e. 
   Gly,  two for the couple     (U,A), i.e.   Gly and Arg and  three for  the couple (C,G),  i.e. Gly,  Val and Leu.  

\subsection{Invertebrates}

In Table \ref{r-inv} we report the  value of the correlation matrix for the six independent couples of probabilities. 
In Table \ref{rid.var.inv} we report the mean value and the standard
deviation of the probability of usage of the codons $XZN$ $(XZ = NC, CU,
GU, CG, GG)$ computed over all biological species given in Table
\ref{tabledata2}. It can be remarked that this probability shows a rather
large spread which is surprisingly reduced when one makes the sum (compare
Tables \ref{rid.var.inv} and \ref{tablestat-2}).
\begin{table}[htbp]
\begin{tabular}{|c|r|r|r|r|r|r|r|r|r|}
\hline
$r_{XY}$&$r_{CA}$&$r_{UG}$&&$r_{UC}$&$r_{AG}$&&$r_{UA}$&$r_{CG}$\\
\hline \hline
P   &   -0,78   &   -0,63   & &  -0,50   &   -0,74   &&   0,20    &   0,52    \\
T   &   -0,85   &   -0,87   & &  -0,74   &   -0,76   &&   0,62    &   0,60 \\
A   &   -0,82   &   -0,79   & &  -0,75   &   -0,68   &&   0,51    &   0,53    \\
S   &   -0,91   &   -0,83   & &  -0,71   &   -0,86   &&   0,55    &   0,79    \\
V   &   -0,78   &   -0,92   & &  -0,66   &   -0,78   &&   0,72    &   0,46    \\
L   &   -0,49   &   -0,92   & &  -0,48   &   -0,66   &&   0,50    &   0,25    \\
R   &   -0,55   &   -0,76   & &  -0,76   &   -0,27   &&   -0,01   &   0,53    \\
G   &   -0,73   &   -0,48   & &  -0,57   &   -0,14   &&   -0,02   &   0,08    \\
\hline
$<r>_{a.a.}$ &   -0,74   &   -0,78   &&   -0,65   &   -0,61   &&   0,38    &   0,47    \\
\hline \hline
\end{tabular}
\centering\caption{Specimen of \textbf{invertebrates}:
value of the correlation coefficient for the 6 independent couples of probabalities for the i 8
quartets.  In the last row the value of the coefficient averaged over the 8 a.a..}
 \label{r-inv}
\end{table} 
From Table \ref{r-inv} we remark anyway an anti-correlation between $P_A$ and $P_C$ and  $P_U$ and $P_G$.  for the a.a. Thr, Ala, Ser and Val.  Moreover, while for the specimens of vertebrates and plants the correlation matrix  shows a pattern of correlation or of anti-correlation, for invertebrates the coefficient $r_{UA}$ shows predomintantly a correlation for almost all the a.a, but a weak anti-correlation for Arg and Gly.  

Si osserva il seguente '' pattern'' di correlazione
\begin{itemize}
    \item P$\rightarrow |r_{CA}|>     |r_{AG}|     >|r_{UG}|      >|r_{CG}|\simeq |r_{UC}|      >|r_{UA}|$
    \item S$\rightarrow |r_{CA}|>     |r_{AG}|     >|r_{UG}|      >|r_{CG}|      >|r_{UC}|      >|r_{UA}|$

    \item T$\rightarrow |r_{CA}|\simeq|r_{UG}|     >|r_{AG}| \simeq|r_{UC}|      >|r_{UA}|\simeq |r_{CG}|$
    \item A$\rightarrow |r_{CA}|>     |r_{UG}|     >|r_{UC}|      >|r_{AG}|      >|r_{CG}|\simeq |r_{UA}|$

    \item G$\rightarrow |r_{CA}|>     |r_{UC}|     >|r_{UG}|      >|r_{AG}|      >|r_{CG}|      >|r_{UA}|$
    \item V$\rightarrow |r_{UG}|>     |r_{CA}|\simeq|r_{AG}|      >|r_{UA}|      >|r_{UC}|      >|r_{CG}|$
    \item L$\rightarrow |r_{UG}|>     |r_{AG}|>     |r_{UA}|\simeq |r_{CA}|\simeq |r_{UC}|      >|r_{CG}|$

    \item R$\rightarrow |r_{UG}|>     |r_{UC}|     >|r_{CA}|\simeq |r_{CG}|      >|r_{AG}|      >|r_{UA}|$
\end{itemize}
that is \emph{for almost all the quartets} the highest value of the correlation coefficient is {\bf always} between $P_{C}$ and $P_{A}$ or $P_{U}$ and $P_{G}$
(P, T, A, S, G, R, respectively L and R). The lowest value for the correlation coefficient  is  between$P_{C}$  and $P_{G}$ or between $P_{U}$ and $P_{A}$ (P T A S R G,  resp. V,L).

\begin{figure}[htbp]\centering
\includegraphics[width=16cm]{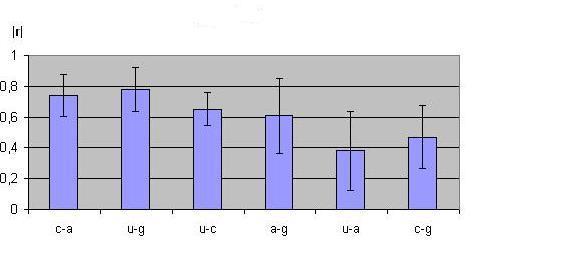}\\
\centering \caption{Specimen of \textbf{invertebrates}: Absolute value of the six highest correlation coefficient  averaged over 8 a.a.. The error is computed by a  standard deviation.}\label{figure/Invcorr.jpg}
\end{figure}

 \begin{figure}[htbp]\centering
\includegraphics[width=16cm]{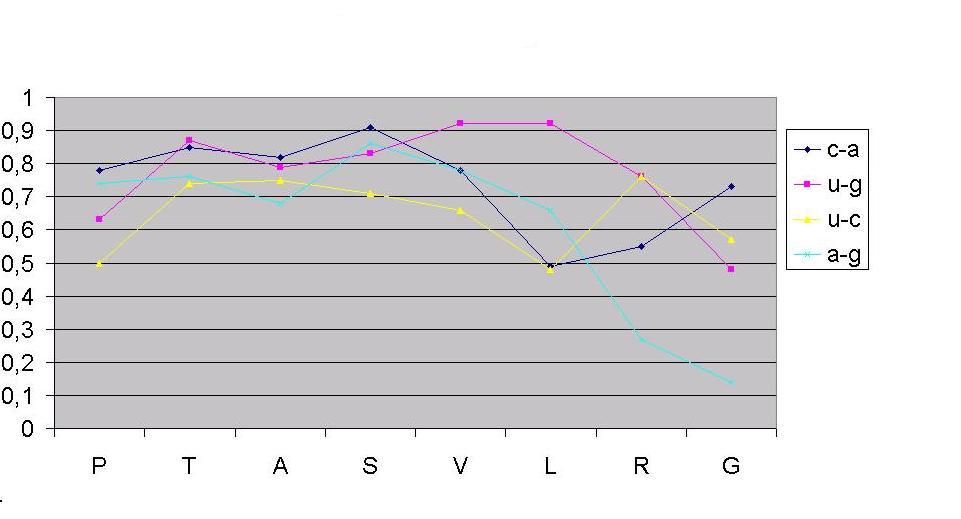}\\
\caption{Specimen of \textbf{invertebrates}: absolute value of the six highest correlation coefficient  averaged over 8 a.a.. correlation coefficient between codon frequency usage for quartets ending in (C,A),(U,G),  (U,C) and (A,G). We do not report the correlations 
 for quartets ending in (U,A) and (C,G), which are in general very low.}
\label{INv.jpg}\
\end{figure}

\subsubsection{Test $\rho=0$ for INVERTEBRATES  n=28}

From eq.(\ref{eq:t}), we get that the critical value of the statistical label  $t$ (95\% confidence level) is
$${t_{c}=t_{5\%}^{n-2}=t_{0,025}^{26}=2,05}$$
\begin{itemize}
    \item   for $t_{r} \in [2,05; -2,05]$ the probability that $\rho=0$ (uncorrelated variables) 
       is 95\%.

    \item   for $t_{r}> 2,05 \quad $or $\quad
    t_{r}<-2,05$  the probability that $\rho=0$ is 5\%  
\end{itemize}

 In Tables \ref{t-inv} and \ref{Int.pln} we report, respectively,  the computed value of the $t$ label for any a.a. and the interval where the values of the correlation coefficient $r$ are included in, at 95 \% confidence level.

\begin{center}
\begin{table}[h]
\begin{tabular} {|c|r|r|r|r|r|r|r|r|r|}
\hline
$INV $&$r_{CA}$&$r_{UG}$&$r_{UC}$&$r_{AG}$&$r_{UA}$&$r_{CG}$\\
\hline \hline
t[P]   &   -6,2    &   -4,1    &   -3,0    &   -5,6           &   \textbf{1,0}   &   3,1 \\
t[T]   &   -8,2    &   -8,9    &   -5,6    &   -6,0           &   4,0            &   3,8 \\
t[A]   &   -7,2    &   -6,6    &   -5,8    &   -4,7           &   3,1            &   3,2 \\
t[S]   &   -10,9   &   -7,7    &   -5,1    &   -8,5           &   3,4            &   6,6 \\
t[V]   &   -6,4    &   -11,8   &   -4,5    &   -6,3           &   5,3            &   2,7 \\
t[L]   &   -2,9    &   -11,8   &   -2,8    &   -4,5           &   2,9            &   \textbf{1,3} \\
t[R]   &   -3,4    &   -6,0    &   -6,0    &   \textbf{-1,4}  &   \textbf{0,0}   &   3,2 \\
t[G]   &   -5,5    &   -2,8    &   -3,6    &   \textbf{-0,7}  &   \textbf{-0,1}  &   \textbf{0,4} \\
\hline
$<t>_{a.a.}$    &   -6,34   &   -7,46   &   -4,55   &   -4,71   &   2,45    &   3,04    \\
\hline \hline
\end{tabular}
\centering\caption{Specimen of \textbf{invertebrates}:
value of the statistical label $t$  for the  6  correlation coefficients for the 
 8 a.a.. In the last row the average value of   $t$.} \label{t-inv}
\end{table}
\end{center}

\begin{table}[htbp]
\begin{tabular}{|c|c|c|c|c|c|c|c|c|c|}
\hline
INV &$r_{CA}$&$r_{UG}$&&     $r_{UC}$ & $r_{AG}$& &     $r_{UA}$&$r_{CG}$\\
\hline \hline
P   &   -0,57   -0,89   &   -0,34   -0,81   &&  -0,16   -0,74    &   -0,51   -0,87   &&   -0,19   0,53    &   0,18    0,75    \\
T   &   -0,70   -0,93   &   -0,74   -0,94   &&   -0,51   -0,87   &   -0,54   -0,88   &&   0,32    0,81    &   0,29    0,8 \\
A   &   -0,64   -0,91   &   -0,59   -0,90   &&   -0,52   -0,88   &   -0,41   -0,84   &&   0,17    0,74    &   0,2 0,75    \\
S   &   -0,81   -0,96   &   -0,66   -0,92   &&   -0,46   -0,86   &   -0,72   -0,93   &&   0,22    0,77    &   0,59    0,9 \\
V   &   -0,57   -0,89   &   -0,83   -0,96   &&   -0,38   -0,83   &   -0,57   -0,89   &&   0,47    0,86    &   0,1 0,71    \\
L   &   -0,14   -0,73   &   -0,83   -0,96   &&   -0,13   -0,72   &   -0,38   -0,83   &&   0,16    0,74    &   -0,14   0,57    \\
R   &   -0,22   -0,77   &   -0,54   -0,88   &&   -0,54   -0,88   &   0,11    -0,58   &&   -0,38   0,36    &   0,2 0,75    \\
G   &   -0,49   -0,87   &   -0,13   -0,72   &&   -0,25   -0,78   &   0,25    -0,49   &&   -0,39   0,36    &   -0,3    0,44    \\
\hline
\end{tabular}
\centering \caption{Confidence Intervals (95\%  confidence level) for the specimen of  
invertebrates n=28} \label{Int.inv}
\end{table}

 \begin{figure}[htbp] \centering
\includegraphics[width=16cm]{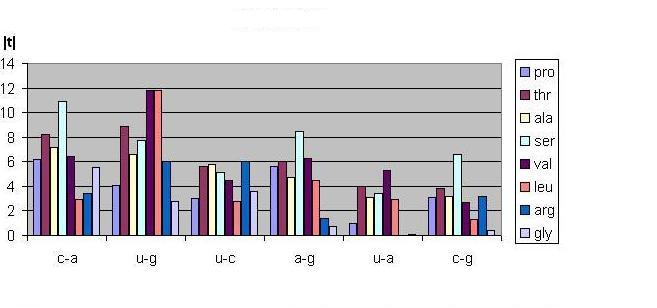}\\
\caption{Value of $|t|$ for the correlation coefficients for the specimen of invertebrates.}
\label{figure/T0inv.jpg} \end{figure}

From  Table \ref{t-inv} we remark, also in this case, that the value of the modulus of the statistical label $t$    computed over the specimen of invertebrates is, in the average, more far from the critical value
   $t_c=2,05$ for the correlation $r_{CA}$ and $r_{UG}$ than for $r_{UC}$ and $r_{AG}$.  From  Table \ref{Int.inv},  one correlation factor can be vaninishing (95\%  confidence level)   for the couple    (A,G), i.e. Arg,  three for the couple (U,A), i.e. Arg, Gly and Pro, and at least two  for the couple     (C,G),  i.e.   Leu and Gly. 

\newpage

\subsection{Mitochondrial code}
 
 In Table \ref{r-mit}  we report 
  the correlations coefficient for the 6 independent couples of probabilities for the  8
quartets for the specimen of \textbf{mitochondria of vertebrates},  see Table \ref{tabledata1m}.
 \begin{table}[htbp]
\begin{tabular}{|c|r|r|r|r|r|r|r|r|r|}
\hline
$r_{XY}$&$r_{CA}$&$r_{UG}$&& $r_{UC}$&$r_{AG}$&& $r_{UA}$&$r_{CG}$\\
\hline \hline
P   &   -0,78   &   0,28    &&   -0,14   &   -0,22   &&   -0,49   &   -0,12   \\
T   &   -0,84   &   0,38    &&   -0,33   &    0,02   &&   -0,23   &   -0,32   \\
A   &   -0,78   &   -0,16   &&   -0,68   &    0,19   &&   0,10    &   -0,21   \\
S   &   -0,83   &   -0,15   &&   -0,49   &   -0,74   &&   -0,05   &   0,63    \\
V   &   -0,59   &   -0,08   &&   -0,44   &   -0,49   &&   -0,41   &   0,26    \\
L   &   -0,34   &   -0,22   &&   -0,64   &   -0,21   &&   -0,48   &   0,22    \\
R   &   -0,70   &   -0,12   &&   -0,18   &   -0,33   &&   -0,32   &   -0,19   \\
G   &   -0,69   &   0,57    &&   -0,41   &   -0,39   &&   -0,31   &   -0,29   \\
\hline
$<r>_{a.a.}$ &  -0,69  &   0,06    &&   -0,41   &   -0,27   &&   -0,27    &   -0,02    \\
\hline \hline  
\end{tabular}
\centering \caption{Mitochondria vertebrates:value of the correlation coefficient for the 6 independent couples of probabalities for the 8 quartets.  In the last row the value of the coefficient averaged over the 8 a.a..} 
\label{r-mit}
\end{table}

We remark that
\begin{itemize}
\item only the coefficient $r_{CA}$ for Pro, The, Ala and Ser 
 has a relatively high value ($r \sim 0,8$), but the coefficient $r_{UG}$ for the complementary couple does not show this property. The other values do indicate the absence of correlation. 
 \item the c.u.f. for mitochondria is more asymmetric that for eukaryotes. It seems the codons
 $XYA$  and $XYC$ are most used while the remaining codons, in particular $XYC$, are very poorly used.
 \end{itemize}
 
\begin{figure}[htbp]\centering
\includegraphics[width=16cm]{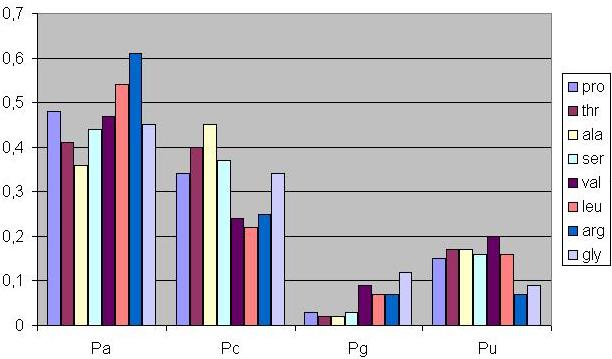}\\
\caption{Probability of the four codons in the 8 a.a. for the mitochondrial code.}\label{Pmit.jpg}\
\end{figure}

  In Fig.(\ref{Pmit.jpg}) one can remark that the codon frequency distribution on the 4 codons  is more  {\bf asymmetric} than in the analogous figures for eukaryotic species
The mitochondrial genetic information seems to be preferably encoded by codons  XZA and XZC, the two remaining codons, in particular XZG,  are for the most part  little used or
resolutely  suppressed. In particular in Fig.(\ref{Pmit.jpg}) we remark,in correspondence of the quartets  P,T,A,S ($r_{CA} \sim 0,8$), $P(XZG) \sim 0,05$ and $P(XZU)\sim 0,15$, so$P_{A+C}\approx1-0,20=0,80$, as it is obtained by averaging over the 20 species.

\subsubsection{Test $\rho=0$ for Mitochondres  n=20}
 
From eq.(\ref{eq:t}), we get that the critical value of the statistical label  $t$ (95\% confidence level) is
$${t_{c}=t_{5\%}^{n-2}=t_{0,025}^{18}=2,1}$$

\begin{itemize}
    \item   for $t_{r} \in [2,1; -2,1]$ the probability that $\rho=0$ (uncorrelated variables) 
       is 95\%.

    \item   for $t_{r}> 2,1 \quad $ or $\quad
    t_{r}<-2,1$  the probability that $\rho=0$ is 5\%  
\end{itemize}

 In the Table  \ref{t-mit} we report  the computed value of the $t$ label for any a.a.  and we indicate in bold the values of the correlation coefficients for whch it is plausible to take the value $\rho =0$ (uncorrelated at 95\% confidence level).
\begin{table}[htbp]
\begin{tabular}{|c|r|r|r|r|r|r|r|}
\hline
MIT &$r_{CA}$&$r_{UG}$&$r_{UC}$&$r_{AG}$&$r_{UA}$&$r_{CG}$\\
\hline \hline
t[P]   &   -5,29            &   \textbf{1,24}  &   \textbf{-0,60} &   \textbf{-0,96}&   -2,38           &   \textbf{-0,51}   \\
t[T]   &   -6,57            &   \textbf{1,74}  &   \textbf{-1,48} &    \textbf{0,08}&   \textbf{-1,00}  &   \textbf{-1,43}   \\
t[A]   &   -5,29            &   \textbf{-0,69} &   -3,93          &   \textbf{0,82} &   \textbf{0,43}   &   \textbf{-0,91}   \\
t[S]   &   -6,31            &   \textbf{-0,64} &   -2,38          &   -4,67         &   \textbf{-0,21}  &   3,44    \\
t[V]   &   -3,10            &   \textbf{-0,34} &  \textbf{ -2,08}      &   -2,38         &    \textbf{-1,91} &   \textbf{1,14}    \\
t[L]   &   \textbf{-1,53}   &   \textbf{-0,96} &   -3,53          &   \textbf{-0,91}&   -2,32           &   \textbf{0,96}    \\
t[R]  &   -4,16            &   \textbf{-0,51} &   \textbf{-0,78} &   \textbf{-1,48}&   \textbf{-1,43}  &   \textbf{-0,82}   \\
t[G]   &   -4,04            &   2,94           &   -\textbf{1,91} &   \textbf{-1,8} &   \textbf{-1,38}  &   \textbf{-1,29}   \\
\hline
\end{tabular}
\centering\caption{Mitochondria vertebrates:
value of the statistical label $t$  for the  6  correlation coefficients for the 
 8 a.a..} \label{t-mit}
\end{table}

A general survey of Table \ref{t-mit} confirms that the absence of any correlation is consistent with the obtained results.

\begin{figure}[htbp]\centering\footnotesize
\includegraphics[width=16cm]{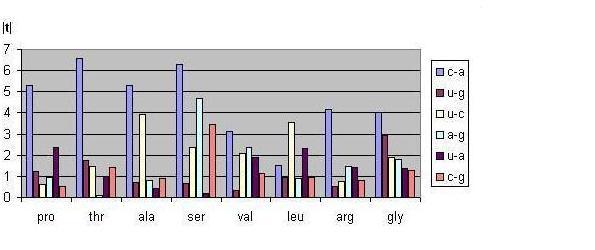}

\caption{Mitochondrial specimen: diagram of  $|t|$.}
\end{figure}
Let us remark that the  $|t|$ diagram for the mitochondrial specimen shows that most of the value of  $|t|$ are below the critical value $t_{c}=2,1$.

\section{Correlation between amino-acids}

In \citep{FSS1}, \citep{CFSS} it was remarked that the values of the ratio
\be
B_{NN'} = \frac{P_{XYN}}{P_{X'Y'N'}}
\ee
were independent on the considered biological species and, for any biological species, were very close each other for $XY, X'Y'  \in  \{CU, GU\}$ or $XY, X'Y' \in  \{CC, UC, GC, AC\}$, i.e. for the quartets encoding, respectively, Leu, Val and Pro, Ser, Ala, Thr. This feature suggests the possible existence of a correlation between  $P_{XYN}$ and $P_{X'Y'N}$ for $XY, X'Y'$ belonging to one of the two above specified sets. In order to search for correlation we computed the $28 \times 28$ correlation matrix $r_{XYN,X'Y'N} \equiv  r_{P_{XYN},P_{X'Y'N}}$ for the 8-dinucleotides corresponding to  quartets. We do not report here the whole matrix, but the following interesting correlation pattern comes out
\begin{table}[htbp]
\begin{tabular}{|c|c||c|c|c|c|} \hline
 a.a. & XY-X'Y' & $r_{AA}$  & $r_{CC}$ & $r_{GG}$ &$r_{UU}$ \\
\hline
 Ala-Pro  & GC-CC &  0,86    &   0,93    &   0,93    &   0,81        \\
 Ala-Ser  &  GC-UC &  0,86    &   0,93    &   0,82    &   0,81       \\
 Ala-Thr &   GC-AC & 0,91   &   0,93    &   0,93    &   0,94        \\
   Pro-Ser &  CC-UC &  0,93    &   0,90    &   0,87    &   0,91     \\
 Pro-Thr   & CC-AC &  0.83    &   0,91    &   0.93   &   0.74    \\
 Ser-Thr  &  UC-AC &0.88    &   0,94    &   0.90   &   0.80     \\
\hline\hline
 Leu-Val  & CU-GU & 0.85    &   0,82    &   -0.70   &   0.96     \\
\hline
\end{tabular}
\centering \caption{Correlation between probabilities for codons in quartets encoding different amino acids computed over the whole eukariotes specimen $n=92$. } \label{r-aa}
\end{table}

We should say that, in the 406 independent entries {\footnote{This number is computed assuming the four $P_{XYN}$ as independent variables.}} of the correlation matrix for the 8 quartets,  several values larger than, let us say, 0.75 do appear, but only for the a.a. listed in Table \ref{r-aa} all the entries are, with exception of $r_{CUG,GUG}$, which further is the only negative values, equal or  larger  than 0.75. In order to get an evaluation of the effects of the statistics of the number of codons, as well as to increase the statistics of the analysed specimen, we have computed this correlation matrix over a specimen of $n=218$ biological species with a codons statistics larger than 30.000 codons. For the 6 a.a. Ala, Pro, Ser, Thr, Val and Leu, we extract the following  correlation pattern reported in Table \ref{r-aa-all}
\begin{table}[htbp]
\begin{tabular}{|c|c|||c|c|c|c|} \hline
 a.a. & XY-X'Y' &$r_{AA}$  & $r_{CC}$ & $r_{GG}$ &$r_{UU}$ \\
\hline
 Ala-Pro  & GC-CC &   0,73    &   0,89    &   0,91    &   0,74        \\
 Ala-Ser  &  GC-UC &  0,86    &   0,85    &   0,79    &   0,80       \\
 Ala-Thr &  GC-AC &  0,77   &   0,87    &   0,91    &   0,85        \\
   Pro-Ser &   CC-UC & 0,82    &   0,82    &   0,85    &   0,82     \\
 Pro-Thr   &  CC-AC &  0.74    &   0,80    &   0.91   &   0.73    \\
 Ser-Thr  & UC-AC &  0.80    &   0,87    &   0.88   &   0.79     \\
\hline\hline
 Leu-Val  &  CU-GU & 0.80    &   0,73    &   -0.61   &   0.94     \\
\hline
\end{tabular}
\centering \caption{Correlation between probabilities for codons in quartets encoding different amino acids computed over  a specimen of $n=218$ species with a codons statistics $\geq$ 30.000.} 
 \label{r-aa-all}
\end{table}

From a comparison of  Table \ref{r-aa} and of Table \ref{r-aa-all} we see that the general pattern is unchanged, even if the values of the correlation coefficients are lower. Looking at these Tables it shows out that the set of eight quartes splits into three subsets
\begin{itemize}
\item a set of 4 a.a.  (Pro, Ser, Ala, Thr) which shows a correlation between the c.u.f. for codon with the same third nucleotide
\item a set of 2 a.a. (Leu, Val) which shows a correlation between the  c.u.f. for codon with the same third nucleotide
\item a set of 2 a.a. (Arg, Gly)  (not reported in Table) with generally uncorrelated c.u.f.
\end{itemize}

This pattern fits well in the crystal basis model of the genetic code. Indeed in the model the codons belonging to the quartets encoding Pro, Ser, Ala, Thr are identified by elements of the same mathematical space; the same happens for the quartets encoding Leu, Va, while the remaining two quartes, encoding for Arg an Gly, do not share this feature.

\section{Pattern of the Shannon entropy}

 The Shannon entropy, defined by
\be
  S = -\sum_{i} \, f_i  \, \ln_{2}  f_i
          \label{eq:shan}
          \ee
  is largely used in biology, mainly to compute  the distance between the observed frequencies   $ f_i $
  of some quantity and the theoretically predicted ones. In \citep{FSS3} it has been  argued that the contribution of each quartet,  in the total exonic region, to the Shannon entropy, defined as:
\begin{equation}
S_{XY} = -\sum_ {N=C,A,U,G} p_{XYN} \, \ln_{2} p_{XYN}   
\label{eq:shannon}
\end{equation}
 should    be independent on the biological species. Pay attention to not confuse  the previous defined probability $P(XYN) \equiv P_N$ with the probability $p_{XYN}$
               with the normalization condition\footnote{The correct normalization should be performed excluding the three stop codons. However the error induced by the assumed normalization is completely negligible at the  level of accuracy of present paper.}  
  \be
   \sum_ {X,Z,N \in C,A,U,G} \, p_{XZN} \, = \, 1
   \ee
     A consequence of this statement is that  biological species with the same percentage,  in the total exonic decoded region, of a given a.a. should have the same Shannon entropy, computed by eq.(\ref{eq:shan}).  As we are dealing with  a.a. encoded by quartets, the a.a. can be also identified  by the first two nucleotides $XY$, so we should have
   \be
  S_{XY} = -\sum_ {N=C,A,U,G} p_{XYN} \, \ln_{2} p_{XYN}  = -\sum_ {N=C,A,U,G} \widehat{p}_{XYN} \, \ln_{2} \widehat{p}_{XYN} \label{eq:shannon2}
  \ee
  where $p_{XYN}$ and $\widehat{p}_{XYN}$ are the codon usage frequencies for the quartet encoding the a.a. specified by the dinucledotide $XY$ for two different  biological species with the condition
  \be
  \sum_ {N=C,A,U,G} p_{XYN} \,  =  \, \sum_ {N=C,A,U,G} \widehat{p}_{XYN}  \,  \equiv \mbox{percentage of  thw a.a. XY in the exonic decoded region}  \label{eq:id}
  \ee 
  In Table \ref{shan} and Table \ref{shan-2} we report, respectively, for a set of two and three different biological species, belonging to the considered specimen,  which have  approximately the same percentage of a given a.a. the computed Shannon entropy. The data confirm largely the independence of the value of the entropy from the values of $p_{XYN} $ peculiar of the considered b.sp..
  However care must be used in concluding that  the calculated Shannon entropy values satisfy eq.(\ref{eq:id}). Indeed one has to take into account that the values of $p_{XYN}$ are quite small (for the quartets $ 5\cdot 10^{-2} > p_{XYN} >   2\cdot 10^{-3}$). So it is natural to wonder if one is really observing a  property of the entropy  or an artefact due  to the smallness of the frequencies. In order to shed light on the question, we have made  some simulations which    confirm the validity of eq.(\ref{eq:id}).
  If one consider, e.g.,  the case of Thr, which appears in the coding region of Xenopus Laevis, Plasmodium  vivax and Plasmodium patents, respectively,  with percentage value 5,270 \%,, 5,267 \%  and 5,279 \%, see Table  \ref{shan-2},   changing of approximately 10 \%   the observed c.u.f., the value of the Shannon entropy  is modified in the second decimal digit.  In the average a change of the c.u.f.   of   the order of a standard deviation  does not affect the third decimal digit. These considerations do not completely solve the above raised question, but  give a cue  the equality of the values of the Shannon entropy in Tables \ref{shan} and  \ref{shan-2} to be an indication of this (unexpected) feature.
  
  From the statement of eq.(\ref{eq:shannon2}) it follows that the knowledge of the c.u.f., typical of a fixed biological species, does not provide further information on the Shannon entropy of the a.a., i.e. the knowledge of the four quantities $p_{XYN}$,  constrained by the condition eq.(\ref{eq:id}) gives the same amount of information of the knowledge of only one quantity $\{XY\} (\equiv $  percentage of  the a.a. XY).  This is  understandable, on general grounds, if the four $p_{XYN}$ are constrained by two further constraints, which may be the correlations, respectively, between $p_{XYC}$ and $p_{XYA}$ and between $p_{XYG}$ and $p_{XYU}$.  That being the case the knowledge of  the four $p_{XYN}$ in fact reduces to the knowledge of only one independent  quantity equivalent to the knowledge of $\{XY\}$.
  So the computation of  the Shannon entropy provides a further  support to the existence of the correlations.


\begin{table}[htbp]
\scriptsize
\centering
\begin{tabular}{rrrrrrrrr}
\hline
a.a. & b.sp. & b.sp. & & \% 1st-col & 2nd-col & & $S_{1}(a.a.)$ & $S_{2}(a.a.)$  \\ \hline\hline
 {\bf Ala} &   {\it 15} &         34 &            & {\it 8,425} &      8,421 &            & {\it 0,467} &      0,464 \\
           &    {\it 1} &         26 &            & {\it 7,166} &      7,167 &            & {\it 0,41} &      0,398 \\
           &         23 &         19 &            &      6,866 &      6,868 &            &      0,392 &      0,385 \\
           &   {\it 23} &         29 &            & {\it 5,619} &      5,611 &            & {\it 0,329} &      0,339 \\
 {\bf Arg} &         19 &         38 &            &      3,368 &      3,369 &            &      0,228 &      0,228 \\  \hline
           &         26 &    {\it 5} &            &      3,246 & {\it 3,247} &            &      0,221 & {\it 0,222} \\
           &         14 &    {\it 5} &            &      3,245 & {\it 3,247} &            &      0,223 & {\it 0,222} \\
           &         12 &   {\it 15} &            &      3,133 & {\it 3,134} &            &      0,217 & {\it 0,215} \\
           &         11 &         17 &            &      3,085 &      3,092 &            &      0,215 &      0,213 \\
           &          3 &          6 &            &      3,057 &      3,066 &            &      0,213 &      0,212 \\
           &   {\it 13} &         29 &            & {\it 1,377} &       1,38 &            & {\it 0,111} &       0,11 \\  \hline
 {\bf Gly} &         25 &         23 &            &      7,835 &      7,835 &            &       0,44 &       0,44 \\
           &    {\it 3} &         38 &            & {\it 7,311} &      7,299 &            & {\it 0,415} &      0,415 \\
           &          4 &         28 &            &      7,199 &      7,195 &            &      0,416 &       0,41 \\
           &    {\it 9} &         26 &            & {\it 7,162} &      7,164 &            & {\it 0,391} &      0,404 \\
           &          8 &         35 &            &      6,393 &      6,397 &            &       0,38 &      0,365 \\  \hline
           &   {\it 18} &         27 &            & {\it 5,449} &      5,442 &            & {\it 0,329} &      0,323 \\
 {\bf Leu} &    {\it 2} &         17 &            & {\it 7,496} &      7,499 &            & {\it 0,396} &       0,42 \\
           &         13 &         38 &            &      7,424 &      7,426 &            &      0,406 &      0,417 \\
           &   {\it 15} &          2 &            & {\it 7,247} &      7,249 &            & {\it 0,413} &      0,411 \\
           &          1 &          3 &            &      7,068 &      7,064 &            &       0,39 &      0,401 \\
           &    {\it 4} &         22 &            & {\it 5,607} &      5,608 &            & {\it 0,34} &      0,339 \\  \hline
 {\bf Pro} &         16 &    {\it 5} &            &      6,126 & {\it 6,125} &            &      0,364 & {\it 0,364} \\
    {\bf } &         13 &    {\it 5} &            &      6,123 & {\it 6,125} &            &      0,364 & {\it 0,364} \\
           &          5 &   {\it 24} &            &      5,714 & {\it 5,715} &            &      0,348 & {\it 0,346} \\
           &         23 &   {\it 12} &            &      5,491 & {\it 5,497} &            &      0,334 & {\it 0,334} \\
           &   {\it 26} &         37 &            & {\it 5,44} &      5,439 &            & {\it 0,336} &      0,326 \\
           &          4 &         18 &            &      5,149 &       5,15 &            &      0,317 &      0,323 \\
           &    {\it 1} &         25 &            & {\it 5,028} &      5,032 &            & {\it 0,315} &      0,317 \\
           &    {\it 4} &          6 &            & {\it 4,901} &      4,909 &            & {\it 0,295} &      0,299 \\  \hline
 {\bf Ser} &   {\it 18} &          9 &            & {\it 6,077} &      6,067 &            & {\it 0,366} &       0,36 \\
           &          7 &    {\it 8} &            &      5,109 & {\it 5,109} &            &      0,314 & {\it 0,315} \\
           &   {\it 24} &         27 &            & {\it 5,09} &      5,098 &            & {\it 0,319} &      0,302 \\
           &    {\it 3} &         15 &            & {\it 5,041} &      5,041 &            & {\it 0,318} &      0,311 \\
           &          2 &   {\it 27} &            &      5,033 & {\it 5,027} &            &      0,313 & {\it 0,317} \\
           &         12 &         33 &            &          5 &      5,003 &            &      0,309 &      0,315 \\
           &         18 &   {\it 13} &            &      4,984 & {\it 4,989} &            &      0,306 & {\it 0,314} \\
           &         15 &   {\it 14} &            &      4,747 & {\it 4,744} &            &      0,297 & {\it 0,267} \\
           &         15 &   {\it 19} &            &      4,747 & {\it 4,744} &            &      0,297 & {\it 0,287} \\
           &         26 &   {\it 17} &            &      4,368 & {\it 4,369} &            &      0,278 & {\it 0,284} \\ \hline
 {\bf Thr} &    {\it 3} &         36 &            & {\it 5,826} &      5,822 &            & {\it 0,355} &      0,342 \\
           &    {\it 2} &          9 &            & {\it 5,774} &      5,774 &            & {\it 0,341} &      0,345 \\
           &   {\it 16} &         35 &            & {\it 5,774} &       5,77 &            & {\it 0,344} &      0,348 \\
           &         11 &   {\it 26} &            &      5,684 & {\it 5,683} &            &      0,342 & {\it 0,348} \\
           &          7 &   {\it 26} &            &      5,687 & {\it 5,683} &            &      0,347 & {\it 0,348} \\
           &         11 &          7 &            &      5,684 &      5,687 &            &      0,342 &      0,347 \\
           &          1 &         26 &            &      5,479 &      5,474 &            &      0,331 &      0,329 \\
           &         10 &         26 &            &      5,479 &      5,474 &            &      0,331 &      0,329 \\
           &          7 &    {\it 1} &            &      5,389 & {\it 5,384} &            &      0,327 & {\it 0,331} \\
           &         16 &         16 &            &      5,309 &      5,307 &            &      0,326 &      0,325 \\
           &         17 &         10 &            &      5,289 &      5,282 &            &      0,324 &       0,31 \\
           &         24 &         25 &            &      5,278 &      5,279 &            &      0,324 &      0,329 \\
           &         25 &   {\it 22} &            &       5,27 & {\it 5,267} &            &       0,32 & {\it 0,328} \\
           &   {\it 25} &         18 &            & {\it 5,109} &       5,11 &            & {\it 0,297} &      0,317 \\
           &   {\it 19} &         15 &            & {\it 5,059} &      5,056 &            & {\it 0,298} &      0,312 \\ \hline
 {\bf Val} &   {\it 17} &         25 &            & {\it 7,224} &      7,228 &            & {\it 0,413} &      0,408 \\
           &   {\it 16} &         17 &            & {\it 7,011} &      7,017 &            & {\it 0,384} &      0,397 \\
           &          1 &         10 &            &        6,9 &      6,925 &            &      0,392 &      0,357 \\
           &         19 &         18 &            &      6,715 &      6,719 &            &      0,376 &      0,388 \\
           &          4 &         37 &            &      6,701 &      6,705 &            &      0,381 &      0,382 \\
           &         25 &    {\it 2} &            &      6,601 & {\it 6,607} &            &      0,369 & {\it 0,378} \\
           &         13 &         28 &            &      6,017 &      6,016 &            &      0,352 &      0,354 \\
           &         22 &         12 &            &      6,091 &       6,08 &            &      0,351 &      0,363 \\
           &         17 &         21 &            &      5,954 &      5,958 &            &      0,355 &      0,352 \\ \hline
\end{tabular} 
 \caption{\footnotesize{In the first column it is indicated the a.a.. In the second  and third column the number identifies the species denoted for vertebrates (bold font, {\bf x}),  for plants (normal font, x) and for invertebrates (italic font, {\it x}). In the next two columns the percentage of the a.a.  and  In the last two  columns the computed Shamnon entropy relative to the  a.a.. for the biological species specified in the 2nd anf 3rd column.}} \label{shan}
\end{table}

 \begin{table}[htbp]
\footnotesize
\centering
\begin{tabular}{rrrrrrrrrr}
 \hline  
 a.a. &  VRT & INV  &    PLN   & \% a.a. V &  \% a.a. I & \% a.a. I   &  $ S^{a.a.}$ V & $ S^{a.a.}$ I & 
 $S^{a.a.}$  P  \\  \hline \hline   
 {\bf Ala} &  3 & 1    &      26 &              7,156  &  7,166     &      7,167       &  0,41 &  0,41 & 0,398     \\
           &  13 & 22    &      15 &              7,025 &  7,084     &      7,115       &  0,40 &  0,407 & 0,402     \\
           &  9 & 18    &      32 &              6,767 &  6,74     &      6,808     &  0,388 &  0,396 & 0,382     \\
           &  7 & 10    &      30 &             7,344 &  6,336     &      6,262       &  0,369 &  0,367 & 0,361     \\ \hline
 {\bf Arg} &    19      & 3    &      38             &  3,368 &  3.347     &   3,369       &  0,228 &  0,229 & 0,222     \\
           &  16 & 4    &      18 &               3,286 &  3,313     &   3,273       &  0,225 &  0,225 & 0,222     \\
           &  25 & 5    &      30 &               3,264 &  3,247     &   3,267       &  0,224 &  0,222 & 0,219  \\
           & 12 & 15    &      17 &               3,133 &  3,134     &   3,092       &  0,217 &  0,215 & 0,213     \\
           & 22 & 24    &      34 &               2,841 &     2,871   &   2,806       &  0,201 &  0,203 & 0,196     \\\hline
 {\bf Gly} &  21 & 28    &      20             &  7,159 &  7,160    &      7,165      &  0,412 &  0,412 & 0,402     \\
           &  20 & 28    &      12 &              7,139 &  7,160     &      7,130       &  0,411 &  0,412 & 0,411     \\
             &  26 & 10    &      22 &              6,842 &  6,821     &      6,807       &  0,397 &  0,384 & 0,394     \\
              &  24 & 15    &      22 &              6,713 &  6,739    &      6,807       &  0,389 &  0,394 & 0,394     \\
               &  5 & 27    &      22        &       6,702 &  6,701     &      6,807       &  0,392 &  0,393 & 0,394     \\
                &  18 & 14    &      1             &  6,609 &  6,511     &      6,581       &  0,388 &  0,332 & 0,382     \\
                 &  14 & 13    &      1 &              6,548 &  6,505     &      6,583       &  0,386 &  0,374 & 0,382     \\
                  &  3 & 1    &      11 &              6,457 &  6,497     &      6,474       &  0,383 &  0,377 & 0,378     \\
                   &  17 & 23    &      2 &              6,038 &  5,993     &     5,937       &  0,364 &  0,345 & 0,355     \\\hline
 {\bf Leu}  &  17 & 2    &      17 &              7,506 &  7,496    &      7,499      &  0,417 &  0,396 & 0,42     \\
           &  5 & 9    &     37 &              7,443 &  7,453     &      7,459      &  0,401 &  0,382 & 0,41     \\
             &  8 & 8    &      28 &              6,876 &  6,93     &      6,943       &  0,393 &  0,383 & 0,391     \\
  {\bf Pro}&  25 & 16    &    35 &             5,905 &   5,885     &      5,863       &    0,354 &  0,348 & 0,356  \\
          & 5 & 24   &    38 &             5,714 &   5,715     &      5,554       &    0,348 &  0,346 & 0,343  \\
           &  1 & 4    &    19 &             4,894 &  4,901    &    4,909       &    0,308 &  0,297 & 0,299  \\ \hline
 {\bf Ser}  &  13 & 11    &    22             & 5,359 &   5,419     &      5,409       &    0,274 &  0,325 & 0,328  \\
          & 7 & 8   &   25 &             5,109 &   5,109     &      5,126       &    0,314 &  0,315 & 0,322  \\
           &  5 & 24    &   27 &              5,105 &   5,09     &      5,098       &    0,313 &  0,319 & 0,302   \\
            &  2 & 3    &   15 &              5,033 &   5,041     &      5,041       &    0,313 &  0,318 & 0,311   \\
            &  2 & 27    &   15 &              5,033 &   5,027     &      5,041       &    0,313 &  0,317 & 0,311   \\
            &  12 & 27    &   33 &              5 &   5,027     &      5,003       &    0,309 &  0,314 & 0,315   \\
              &  17 & 25    &   34 &             4,917 &  4,875    &    4,888       &    0,304 &  0,267 & 0,309  \\
               &  15 & 14    &   17 &             4,747 &  4,744    &    4,736       &    0,297 &  0,287 & 0,301  \\
                &  24 & 9    &   18 &             4,292 &  4,169    &    4,239       &    0,274 &  0,265 & 0,276  \\ \hline
 {\bf Thr}  &  13 & 23    &    21 &             5,982 &   6,018     &      6,03       &    0,357 &  0,346 & 0,359  \\
          & 11 & 26   &   7 &             5,684 &   5,683     &      5,687       &    0,342 &  0,348 & 0,347  \\
           &  7 & 1    &   24 &              5,389 &   5,384     &      5,411       &    0,327 &  0,331 & 0,331   \  \\
            &  24 & 22    &   25 &              5,278 &   5,267     &      5,279       &    0,324 &  0,328 & 0,329   \\
            &  25 & 22    &   25 &              5,27 &   5,267     &      5,279       &    0,324 &  0,328 & 0,329   \\ \hline
 {\bf Val}   &  1 & 16    &      10 &              6,9 &  7,011     &      6,925       &  0,392 &  0,384 & 0,357     \\
                 &  25 & 2    &      22 &              6,601 &  6,607     &      6,604       &  0,369 &  0,378 & 0,384     \\
                &  18 & 14   &      34 &              6,573 &  6,566     &      6,588       &  0,372 &  0,348 & 0,38     \\
                &  16 & 24    &      12 &              6,099 &  6,137     &      6,08       &  0,357 &  0,368 & 0,363     \\
                 &  17 & 8    &    21 &             5,954 &   5,895     &      5,958       &    0,355 &  0,347 & 0,352 \\  \hline
                 \end{tabular}  
                \caption {In the first column                it is indicated the a.a.. In the second, third, fourth column the number identifies the species, respectively, for vertebrates (V), plants (P) and invertebrates (I). In the next three columns and, respectively, in the last three the percentage of the a.a. and the computed Shamnon entropy, for the indicated biological species, columns.} \label{shan-2}
\end{table}

\newpage

\section{Conclusions}

We can summarize the results as it follows
\begin{itemize}
    \item \texttt{Eucaryotes}: for all  a.a.,  the variance reduction for the probabilities  $P_{U+G}$ and $P_{C+A}$  ùis larger than the reduction for the values of
  $P_{U+C}$, $P_{A+G}$, $P_{C+G}$ and $P_{U+G}$ 
while the value of the ratio  $\sigma^2_{mis}/\sigma^2_{th}$ is close to one
 for $P_{C+G}$ and  $P_{U+G}$, 
 see Table \ref{tablestat}.
 
\begin{table}[h]
 \footnotesize
\begin{tabular}{|c||c||c||c||} \hline
Prob.& $\small{<{\sigma_{obs}}^2/
{\sigma_{th}}^2>_{a.a}^{\texttt{VRT}}}$&$\small{<{\sigma_{obs}}^2/
{\sigma_{th}}^2>_{a.a}^{\texttt{INV}}}$&$\small{<{\sigma_{obs}}^2/
{\sigma_{th}}^2>_{a.a}^{\texttt{PLN}}}$\\
\hline
$P_{C+A}$         &   0,14  &   0,28  &   0,21  \\
$P_{U+G}$      &   0,22  &   0,27  &   0,26     \\
$P_{U+C}$       &   0,31  &   0,38  &   0,33  \\
$P_{A+G}$       &   0,42    &   0,44  &   0,50 \\
$P_{U+A}$       &   1,69   &   1,36  &   0,50 \\
$P_{C+G}$       &   1,36   &   1,44  &   1,50  \\
\hline
\end{tabular}
\centering \caption {\footnotesize{Value of the ratio
$\sigma^2_{obs}(X+Y)$ and ${\sigma^2}_{th}(X+Y) \equiv \sigma^2_{obs}(X)+\sigma^2_{obs}(Y)$ averaged over the  8
aminoacids  -  
$<\sigma^{2}/{\sigma_{th}}^2>_{a.a}$ -  for sum of the probabilities.}} \label{tablestat}
\end{table}

  In more detail:
                       \begin{enumerate}
                       \item Vertebrates: from Table \ref{rid.var.ver}
                         the ratio
                       $\sigma^2/{\sigma^2}_{th}$ is smaller  for $C+A$ than  for  $U+G$ for
                         all a.a.,  except   Arg (R), Leu (L) and 
                      Val (V)  for which the contrary is true.\\
                       In the average, for the 8 a.a. encoded by quartets  we find:
                        \begin{enumerate}
                        \item
                       For
                     $P_{C+A}$   for $P_{U+G}$   $\sigma^2 \simeq 0,015-0,25 \; {\sigma^2}_{th}$ .  So, in the average,  the variance reduction is approximately of one order of magnitude. 
                     \item For $P_{U+C}$ and $P_{A+G}$ $\sigma^2 \simeq 0,3-0,40 \; {\sigma^2}_{th}$ 
                     \item  For $P_{U+A}$ and  $P_{C+G}$ $\sigma^2 \simeq 1.4-1.7 \; {\sigma^2}_{th}$.                      \end{enumerate}
                  
                    \item Invertebrates: from Table \ref{rid.var.inv} the ratio
                     $\sigma^2/{\sigma^2}_{th}$  is smaller for  $U+G$ than for $ C+A$ for all
                        a.a  except Ser (S), Pro (P) and Ala (A) or which the contrary is true.
                        \\
                      In the average, for the 8 a.a. encoded by quartets  we find: 
                      \begin{enumerate}
                     \item For $P_{C+A}$ and $P_{U+G}$ $\sigma^2 \simeq 0,3 \; {\sigma^2}_{th}$ f.                     \item For
                    $P_{U+C}$ and $P_{A+G}$,   
$\sigma^2 \simeq 0,40-0,45 \;
                    {\sigma^2}_{th}$ 
                    \item  For $P_{U+A}$ and $P_{C+G}$   $\sigma^2 \simeq 1.4\; {\sigma^2}_{th}$.                   
                     \end{enumerate}
                  
                   \item Piantes: from Table \ref{rid.var.pln}
                   the ratio
                     $\sigma^2/{\sigma^2}_{th}$  is smaller for $C+A$ than for  $U+G$  for all
                     a.a  except Leu (L) and
                    Val (V) for which the contrary is true.\\
                     In the average, for the 8 a.a. encoded by quartets  we find:
                     \begin{enumerate}
                   \item For
                    $P_{C+A}$ and $P_{U+G}$.$\sigma^2 \simeq 0,20-0,25 \; {\sigma^2}_{th}$      
                    \item For $P_{U+C}$ and $P_{A+G}$ $\sigma^2 \simeq
                    0,30-0,50  \, {\sigma^2}_{th}$  
                    \item For  $P_{U+A}$ and $P_{C+G}$   $\sigma^2 \simeq 1.4-15 \; {\sigma^2}_{th}$.  
                    \end{enumerate}
                                  \end{enumerate}
 In conclusion a  pattern comes out  in which, for the three considered specimens
 the variance reduction  for the couples $A,C$ and $U,G$ is  larger than for the other couples, as it can be inferred by    Tables \ref{eukca} and \ref{eukgu} where  the average values of $P_C$, $P_A$,  $P_{C+A}$.  and, respectively, of $P_G$, $P_U$ and  $P_{G+U}$ are reported .
 
 \item \texttt{Mitochondria}  
                 \begin{description}
                        \item The variance reduction seems appeciable only for the couple  $C+A$ for which,  in the average, $\sigma^2/{\sigma^2}_{th}\sim
                        0,3$,  while for the remaining couples $\sigma^2/{\sigma^2}_{th}\geq 0,6$.
                     \end{description}
                     \end{itemize}
 So no clear   cue for the existence of correlation in c.u.f. for Mitochondria comes out from the previous analysis.
 
 In \citep{FSS2} a theoretical correlation matrix has been derived from the sum rule for the sum the probabilities of the synonimous codons in quartes with last nucleotide $C$ and $A$., which we compare, in Table \ref{Cor.Mat.Oss.} with the computed one for the three eukaryotic specimen. The theoretical value are indicated in round brackets in the first column,   $x$  is a not a priori computable entry.
 \begin{table}[htbp]
\begin{tabular}{|c||c|c|c|}
 \hline
 & VRT & PLN &   INV \\
 \hline
 $<r_{CA}>_{a.a. }$   ({\color{red} -1})          &  -0.89 & - 0.83 &  -0.74      \\
 \hline
$<r_{UG}>_{a.a} ({\color{red} -1})  $             & -0.80 & -0.78 & --0.78    \\
 \hline \hline
 $<r_{UC}>_{a.a}$     ({\color{red} -x})        &  -0.76 & -0.69  & -0.65     \\
\hline
$<r_{AG}>_{a.a}$   ({\color{red} -x})        & -0.64 & -0.55 & -0.61   \\
 \hline \hline
  $<r_{UA}>_{a.a}$  ({\color{red} x})         & 0.74 &  0.53 & 0.38      \\
\hline
$<r_{CG}>_{a.a}$   ({\color{red}  x})        & 0.41 & 0.37 & 0.47   \\
 \hline
\end{tabular}\centering \caption{Observed  averaged value for the correlation matrix, in bracket the theoretical value.}\label{Cor.Mat.Oss.}
\end{table}
  
  Table \ref{Cor.Mat.Oss.} shows that the computed values of $<r_{NN'}>$ differ from the theoretical ones by a quantity of the value of 10-15 \%, excepr for  $<r_{CG}>$  and, for the specimen of invertebrates, for $<r_{UA}>$. For $<r_{CG}>$  the anomalous behaviour may be understood from the known suppression of the dinucleotide  $CG$, see for possible explanation of the suppressione \citep{KFL}, \citep{KM}.
In conclusion one   remarks that there is a strong indication for the existence of correlation between  $P_C$ and $P_A$ (respectively$P_G$ and $P_U$), which further is unexpectedly high for the specimen of plants. Indeed while one can argue that vertebrates are "similar" species in the phylogenetic tree, plants are an extremely large domain of species. The correlation is less evident for invertebrates, but one should keep in mind that this is not even a domain in the life world, May be one should look for correlations in suitable subsets of invertebrates. It comes out that the ratio of $\sigma_{P_{C+A}}$ over $ <P_{C+A}>$ is in the range of 0,02-0,05 for vertebrates, 0,11-0,26 fo plants, 0,09-0,20 for invertebrates. These last two results rise serious question on the validity of the sum rule, still in presence of a clear indication of correlation.
 
  There are several directions which seem worthwhile to be investigated:
 \begin{itemize}
 \item further statistical analysis with better statistics and further tests of the reliability of the presence of the correlations
 \item the main reasons for the codon usage bias are believed to be: the mutational bias, the translation efficiency, the selection mechanism and the abundance of specific anticodons in the tRNA. The universal presence of correlations suggests that its origin may lie in some very general mechanisms, possibly related to the structure of the genetic code. It seems worthwhile to search for mutation-selection models able to explain the pattern of the correlation.  Work is in progess in this direction.
  \item further statistical analysis  of the behaviour of the Shannon entropy to further test the validity of eq.(\ref{eq:shannon2}). 
 \end{itemize}
\medskip

Let us emphasize our claim: we have remarked that the sum of the usage
probabilities of two suitably choosen codons is, within a few percent, a
constant independent on the biological species for vertebrates, which well
fits in the framework of the crystal basis model. Of course one can restate
the above results stating the sum of the probability of codon usage $XZC +
XZA$ is not depending on the nature of the biological specie, without any
reference to crystal basis model. However a deeper analysis of Table
\ref{tablePCA} shows that $P_{C+A}$ for Pro, Thr, Ala, Ser and Gly is of
the order of 0.62, for Leu and Val of the order 0.35 while for Arg is of
order 0.52. In the crystal basis model the roots, i.e. the dinucleotide
formed by the first two nucleotides of the first 5 amino acids belong to
the same irrep. (1,1), the roots of Leu and Val belong to the irrep. (0,1),
while the root of Arg belongs to the irrep. (1,0). This is an interesting
result, especially for Pro whose molecule has a different structure than
the others amino acids (Pro has an imino group instead of an amino group).

\medskip

It is natural to wonder what happens for other biological species. The
green plants exhibits roughly the same pattern, but probably a more
reliable analysis has to be performed considering a splitting into
families. For invertebrates, the large number of existing biological
species and the lack of data with sufficient diversity prevents from
applying a similar analysis. The case of bacteriae is rather interesting.
Eubacteriae seem to avoid this pattern of correlations. This may be the
influence of selection effects which may be stronger or effective in
shorter times in less complicated species. For bacteriae the G+C content
varies in a wide range from 25 \% to 75 \%. Hence one can argue that
biological species with large difference in the G+C content exhibit large
difference in the correlation pattern discussed in this paper. However,
using the Genbank data, one finds for eubacteriae no correlation between
the G+C content and the value of the probabilities eq.(\ref{eq:11}).
 
 \vspace{1cm}

 {\bf Acknowledgments} It is a pleasure to thank M. Nicodemi for discussions and useful suggestions.

\newpage

\section{Appendix}

 \subsection{The crystal basis model of the genetic code}

For completeness let us briefly recall the main features of
 the crystal basis model of the genetic code.
  Each codon $XZN$ is described by a
state belonging to an irreducible representation (irrep.), denoted
$(J_{H},J_{V})^\xi$, of the algebra $U_{q}\big( sl(2)_{H} \oplus sl(2)_{V}
\big)$ in the limit $q \to 0$ (so-called crystal basis); $J_{H}$, $J_{V}$
take (half-)integer values and the upper label $\xi$ removes the degeneracy
when the same couple of values of $J_{H}$, $J_{V}$ appears several times.
As can be seen in Table \ref{tablerep} there are for example four
representations $(\half,\half)^\xi$, with $\xi=1,2,3,4$. 
Within a given representation $(J_{H},J_{V})^\xi$ two more quantum numbers
$J_{H,3}$, $J_{V,3}$ are necessary to specify a particular state.
see Table
\ref{tablerep}, which is reported here to make the paper self-consistent.

in the model,  it appears natural to write the  codonusage
probability as a function of the biological species (b.s.), of the
particular amino-acid and of the labels $J_{H}$, $J_{V}$, $J_{H,3}$,
$J_{V,3}$ describing the state $XZN$. Assuming the dependence
of the amino-acid  completely determined by the set of labels $J's$, 
we write
\begin{equation}
\label{eq:2}
P(XZN) = P(b.s.; J_{H}, J_{V}, J_{H,3}, J_{V,3})
\end{equation}
With the further hypothesis that the r.h.s. of eq.
(\ref{eq:2}) can be written as the sum of two contributions: a universal function $\rho$
independent on the biological species and a b.s. depending function
$f_{bs}$, in \citep{FSS3} the following equation has been written
\begin{equation}
\label{eq:3}
P(XZN) = \rho^{XZ}(J_{H}, J_{V}, J_{H,3}, J_{V,3}) \; + \;
f_{bs}^{XZ}(J_{H}, J_{V}, J_{H,3}, J_{V,3})
\end{equation}
Previous  analysis, see  \citep{CFSS}, suggests the
contribution of $f_{bs}$ is not negligible but  smaller than the
one due to $\rho$.  So the following form form was assumed
\begin{equation}
\label{eq:4}
f_{bs}^{XZ}(J_{H}, J_{V}, J_{H,3}, J_{V,3}) \approx F_{bs}^{XZ}(J_{H};J_{H,3})
\, + \, G_{bs}^{XZ}(J_{V};J_{V,3})
\end{equation}
  In the framework of the model and of the above assumptions, the codon usage
frequencies for the quartets Ala, Gly, Pro, Thr and Val and for the quartet
sub-part of the sextets Arg (i.e. the codons of the form CGN), Leu (i.e.
CUN) and Ser (i.e. UCN). was analysed. \\
For Thr, Pro, Ala and Ser  one writes
 using Table \ref{tablerep} 
and
eqs. (\ref{eq:2})-(\ref{eq:4}), with $N = A,C,G,U$,
\begin{equation}
\label{eq:5}
P(NCC) + P(NCA) = \rho_{C+A}^{NC} + F_{bs}^{NC}(\Third;x) +
G_{bs}^{NC}(\Third;y) + F_{bs}^{NC}(\half;x') + G_{bs}^{NC}(\half;y')
\end{equation}
\begin{equation}
\label{eq:6}
P(NCG) + P(NCU) = \rho_{G+U}^{NC} + F_{bs}^{NC}(\Third;x) +
G_{bs}^{NC}(\Third;y) + F_{bs}^{NC}(\half;x') + G_{bs}^{NC}(\half;y')
\end{equation}
where  $\rho_{C+A}^{NC}$ denotes the sum of the contribution of
the universal function (i.e. not depending on the biological species)
$\rho$ relative to $NCC$ and $NCA$, while the labels $x,y,x',y'$ depend on
the nature of the first two nucleotides $NC$, see Table \ref{tablerep} in \citep{FSS3}.  
Using the results of Table \ref{tablerep}, 
One remarks that the
difference between eq.  (\ref{eq:5}) and eq.  (\ref{eq:6}) is a quantity
independent of the biological species,
\begin{equation}
\label{eq:7}
P(NCC) + P(NCA) - P(NCG) - P(NCU) \;=\; \rho_{C+A}^{NC} - \rho_{G+U}^{NC}
\;=\; \mbox{Const.}
\end{equation}
In the same way, considering the cases of Leu, Val, Arg and Gly, we obtain
with $W = C,G$
\begin{eqnarray}
\label{eq:8}
P(WUC) + P(WUA) - P(WUG) - P(WUU) &=& \rho_{C+A}^{WU} - \rho_{G+U}^{WU}
\;=\; \mbox{Const.} \\
\label{eq:9}
P(CGC) + P(CGA) - P(CGG) - P(CGU) &=& \rho_{C+A}^{CG} - \rho_{G+U}^{CG}
\;=\; \mbox{Const.} \\
\label{eq:10}
P(GGC) + P(GGA) - P(GGG) - P(GGU) &=& \rho_{C+A}^{GG} - \rho_{G+U}^{GG}
\;=\; \mbox{Const.}
\end{eqnarray}
Since the probabilities for one quadruplet are normalised to one, from eqs.
(\ref{eq:6})-(\ref{eq:10}) we deduce that for all the eight amino acids the
sum of probabilities of codon usage for codons with last A and C (or U and
G) nucleotide is independent of the biological species, i.e.
\begin{eqnarray}
\label{eq:11}
P(XZC) + P(XZA) = \mbox{Const.} \qquad (XZ = NC, CU, GU, CG, GG)
\end{eqnarray}

Now let us make
 two important remarks. \\
-- If we write for $\rho(J_{H}, J_{V}, J_{H,3}, J_{V,3})$, or equivalently
$\rho_{C+A}^{NC}$, an expression of the type (\ref{eq:4}), i.e. separating
the $H$ from the $V$ dependence, it follows that the r.h.s. of eqs.
(\ref{eq:5}) and (\ref{eq:6}) are equal, and consequently the probabilities
$P(NCC) + P(NCA)$ and $P(NCU) + P(NCG)$ should be equal, which is not
experimentally verified. This means that the coupling term between the $H$
and the $V$ is not negligible for the $\rho(J_{H}, J_{V}, J_{H,3},
J_{V,3})$ function. \\
-- Summing equations (\ref{eq:5}) and (\ref{eq:6}) we deduce that the
expression $F_{bs}^{NC}(\Third;x) + G_{bs}^{NC}(\Third;y) +
F_{bs}^{NC}(\half;x') + G_{bs}^{NC}(\half;y')$ is actually not depending on
the biological species. {From} eqs. (\ref{eq:5}) and (\ref{eq:6}) for
different values of $N$ and for analogous equations for the other four
quartets, we can derive relations between sums of $F_{bs}^{NC}$ and/or
$G_{bs}^{NC}$ functions which are independent of the biological species.

\subsection{Statistical analysis}

In the following we recall, for completeness, the definition of the statistical quantities used in the text:

\begin{itemize}
\item the standard  deviation
\be
\sigma_{P}=\sqrt{\frac{1}{n(n-1)}(n \sum_{i=1}^{n}
P_{i}^{2})-(\sum_{i=1}^{n} P_{i})^{2}} 
\ee
where $P_N$ is defined in eq.(\ref{eq:PN})
\item the adimensional correlation coefficient $ r_{XY}$
 \bea
\label{r} r_{XY}=r_{(X_{1}.....X_{n};Y_{1}.....Y _{n})}=
\frac{Cov(X,Y)}{\sigma_{X}\sigma_{Y}}  \nonumber \\
=\frac{1}{\sigma_{X}\sigma_{Y}}\sum_{i=1}^{n}
\frac{1}{n}(X_{i}-<X>)(Y_{i}-<Y>) \in [-1,1 ]
\label{eq:rXY}
\eea
where $<X>\:(<Y>)$  is the average value of the variable
and $\sigma_{X} \: (\sigma_{Y})$ the corresponding standard deviation.
\end{itemize}

For two independent normally distributed variables $X$ and $Y$ we have
\be
\sigma_{th}^2(X+Y)=\sigma^2(X)+\sigma^2(Y)
\ee

If $X \equiv P_{N}$ and $Y \equiv P_{N'}$ we  expect the value of the standard deviation of the sum of $X+Y$ to be smaller than the sum of the two corresponding standard deviation, due to the normalization condition
\begin{center}
$\sum_{N=A,C,G,U}P_N=P(A)+P(C)+P(G)+P(U)=1$ \;\;\;\; ($P_{N+N} \equiv P_{N}+P_{N'}$)
\end{center}
\be
\sigma^2(P_{N+N'}) < \sigma^2(P_{N})+\sigma^2(P_{N'})
\ee
However we expect the reduction to be approximately equal for any couple  
$P_{N}$ and $P_{N'}$ extracted in the set of the four probabilities 
\{$P(A)$,$P(C$),$P(G)$,$P(U)$\}, while we remark that a reduction of the standard deviation for the sum 
$P_{A}+P_{C}$ and $P_{U}+P_{G}$ larger than for the other couples of probabilities. 
We can summarize the results as it follows
($\sigma_{th}^2(P_{N+N'}) =  \sigma^2(P_{N})+\sigma^2(P_{N'}$))

\medskip

\begin{table}[htbp]
\medskip
\begin{tabular}{|c|c|c|c|c|}
 \hline
 & XZU &XZC &XZA &XZG \\
 \hline
XZU              &  1&-x  &x  &-1   \\
 \hline
XZC              & -x&1 &-1   &x    \\
 \hline
XZA              &  x&-1  &1  &-x   \\
\hline
XZG              & -1& x& -x &  1   \\
 \hline
\end{tabular}
 \centering  \caption{Correlation matrix for the codon probability from the sum rules (XZ = AC, CC, GC, UC, CU, GU, CG, GG).}
\label{Cor.Mat}

\end{table}


  \begin{figure}[htp]\centering
\includegraphics[width=10cm]{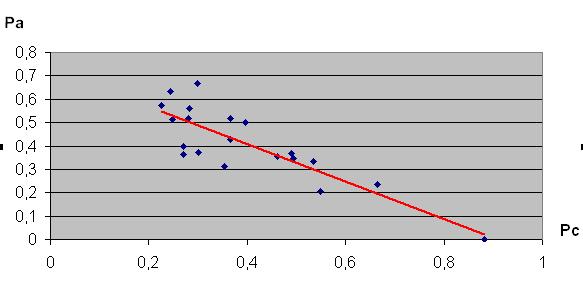}\\
\includegraphics[width=10cm]{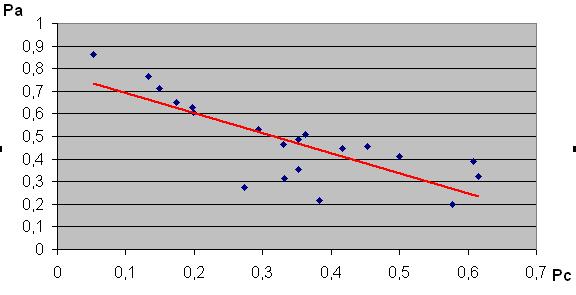}\\
\includegraphics[width=10cm]{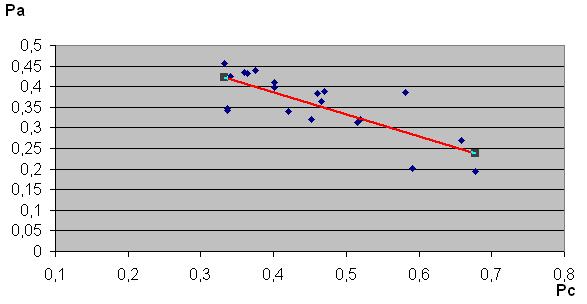}\\
\includegraphics[width=10cm]{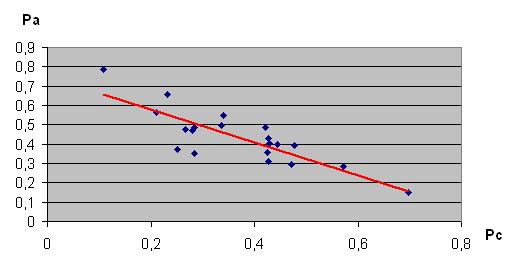}\\
\caption{\footnotesize{Grafico di correlazione per le frequenze
d'uso dei codoni, misurate sul campione di mitocondri, terminanti
per A e C dei quartetti (dall'alto in basso) ACN (Treonina), CCN
(Proplina), GCN (Alalanina), UCN
(Serina)}} \label{CA.mit}
\end{figure}


\begin{figure}[htp]\centering
\includegraphics[width=10cm]{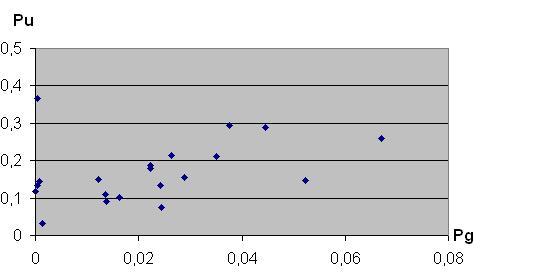}\\
\includegraphics[width=10cm]{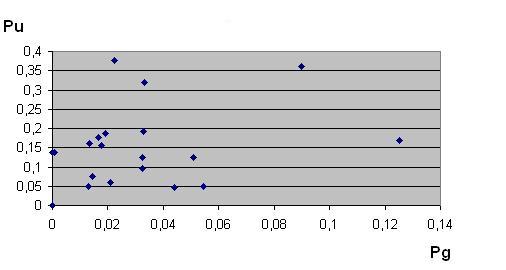}\\
\includegraphics[width=10cm]{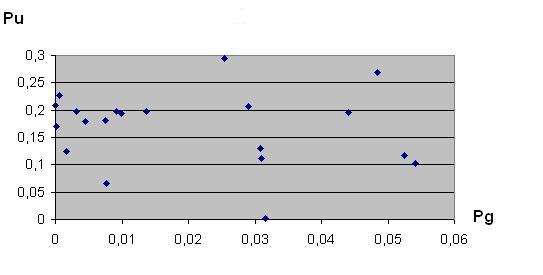}\\
\includegraphics[width=10cm]{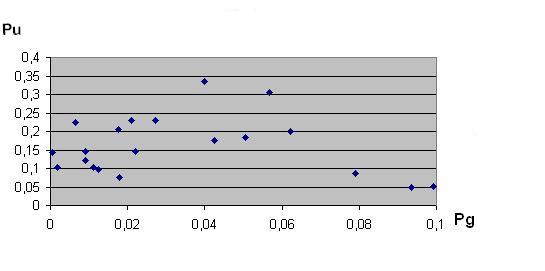}\\
\caption{\footnotesize{Graphic of the codon usage frequency $P_U$ versus $P_G$ for species belonging to the mitochondrial specimen. From top to bottom
 Thr, Pro, Ala and Ser CC}}
 \label{UG.mit}
\end{figure}

\newpage

\begin{table}[htbp]\footnotesize
\caption{The eukaryotic or standard code code. The upper label denotes
different irreducible representations. }
\label{tablerep}
\footnotesize
\begin{center}
\begin{tabular}{|cc|cc|rr|cc|cc|rr|}
\hline
codon & amino acid & $J_{H}$ & $J_{V}$ & $J_{3,H}$ & $J_{3,V}$& codon &
amino acid & $J_{H}$ & $J_{V}$ & $J_{H,3}$ & $J_{V,3}$ \\
\hline
CCC & Pro P & $3/2$ & $3/2$ & $3/2$ & $3/2$ & UCC & Ser S & $3/2$ & $3/2$ &
$1/2$ & $3/2$ \\
CCU & Pro P & $(1/2$ & $3/2)^1$ & $1/2$ & $3/2$ & UCU & Ser S & $(1/2$ &
$3/2)^1$ & $-1/2$ & $3/2$ \\
CCG & Pro P & $(3/2$ & $1/2)^1$ & $3/2$ & $1/2$ & UCG & Ser S & $(3/2$ &
$1/2)^1$ & $1/2$ & $1/2$ \\
CCA & Pro P & $(1/2$ & $1/2)^1$ & $1/2$ & $1/2$ & UCA & Ser S & $(1/2$ &
$1/2)^1$ & $-1/2$ & $1/2$ \\[1mm] \hline
CUC & Leu L & $(1/2$ & $3/2)^2$ & $1/2$ & $3/2$ & UUC & Phe F & $3/2$ &
$3/2$ & $-1/2$ & $3/2$ \\
CUU & Leu L & $(1/2$ & $3/2)^2$ & $-1/2$ & $3/2$ & UUU & Phe F & $3/2$ &
$3/2$ & $-3/2$ & $3/2$ \\
CUG & Leu L & $(1/2$ & $1/2)^3$ & $1/2$ & $1/2$ & UUG & Leu L & $(3/2$ &
$1/2)^1$ & $-1/2$ & $1/2$ \\
CUA & Leu L & $(1/2$ & $1/2)^3$ & $-1/2$ & $1/2$ & UUA & Leu L & $(3/2$ &
$1/2)^1$ & $-3/2$ & $1/2$ \\[1mm] \hline
CGC & Arg R & $(3/2$ & $1/2)^2$ & $3/2$ & $1/2$ & UGC & Cys C & $(3/2$ &
$1/2)^2$ & $1/2$ & $1/2$ \\
CGU & Arg R & $(1/2$ & $1/2)^2$ & $1/2$ & $1/2$ & UGU & Cys C & $(1/2$ &
$1/2)^2$ & $-1/2$ & $1/2$ \\
CGG & Arg R & $(3/2$ & $1/2)^2$ & $3/2$ & $-1/2$ & UGG & Trp W & $(3/2$ &
$1/2)^2$ & $1/2$ & $-1/2$ \\
CGA & Arg R & $(1/2$ & $1/2)^2$ & $1/2$ & $-1/2$ & UGA & Ter & $(1/2$ &
$1/2)^2$ & $-1/2$ & $-1/2$ \\[1mm] \hline
CAC & His H & $(1/2$ & $1/2)^4$ & $1/2$ & $1/2$ & UAC & Tyr Y & $(3/2$ &
$1/2)^2$ & $-1/2$ & $1/2$ \\
CAU & His H & $(1/2$ & $1/2)^4$ & $-1/2$ & $1/2$ & UAU & Tyr Y & $(3/2$ &
$1/2)^2$ & $-3/2$ & $1/2$ \\
CAG & Gln Q & $(1/2$ & $1/2)^4$ & $1/2$ & $-1/2$ & UAG & Ter & $(3/2$ &
$1/2)^2$ & $-1/2$ & $-1/2$ \\
CAA & Gln Q & $(1/2$ & $1/2)^4$ & $-1/2$ & $-1/2$ & UAA & Ter & $(3/2$ &
$1/2)^2$ & $-3/2$ & $-1/2$ \\[1mm] \hline
GCC & Ala A & $3/2$ & $3/2$ & $3/2$ & $1/2$ & ACC & Thr T & $3/2$ & $3/2$ &
$1/2$ & $1/2$ \\
GCU & Ala A & $(1/2$ & $3/2)^1$ & $1/2$ & $1/2$ & ACU & Thr T & $(1/2$ &
$3/2)^1$ & $-1/2$ & $1/2$ \\
GCG & Ala A & $(3/2$ & $1/2)^1$ & $3/2$ & $-1/2$ & ACG & Thr T & $(3/2$ &
$1/2)^1$ & $1/2$ & $-1/2$ \\
GCA & Ala A & $(1/2$ & $1/2)^1$ & $1/2$ & $-1/2$ & ACA & Thr T & $(1/2$ &
$1/2)^1$ & $-1/2$ & $-1/2$ \\[1mm] \hline
GUC & Val V & $(1/2$ & $3/2)^2$ & $1/2$ & $1/2$ & AUC & Ile I & $3/2$ &
$3/2$ & $-1/2$ & $1/2$ \\
GUU & Val V & $(1/2$ & $3/2)^2$ & $-1/2$ & $1/2$ & AUU & Ile I & $3/2$ &
$3/2$ & $-3/2$ & $1/2$ \\
GUG & Val V & $(1/2$ & $1/2)^3$ & $1/2$ & $-1/2$ & AUG & Met M & $(3/2$ &
$1/2)^1$ & $-1/2$ & $-1/2$ \\
GUA & Val V & $(1/2$ & $1/2)^3$ & $-1/2$ & $-1/2$ & AUA & Ile I & $(3/2$ &
$1/2)^1$ & $-3/2$ & $-1/2$ \\[1mm] \hline
GGC & Gly G & $3/2$ & $3/2$ & $3/2$ & $-1/2$ & AGC & Ser S & $3/2$ & $3/2$
& $1/2$ & $-1/2$ \\
GGU & Gly G & $(1/2$ & $3/2)^1$ & $1/2$ & $-1/2$ & AGU & Ser S & $(1/2$ &
$3/2)^1$ & $-1/2$ & $-1/2$ \\
GGG & Gly G & $3/2$ & $3/2$ & $3/2$ & $-3/2$ & AGG & Arg R & $3/2$ & $3/2$
& $1/2$ & $-3/2$ \\
GGA & Gly G & $(1/2$ & $3/2)^1$ & $1/2$ & $-3/2$ & AGA & Arg R & $(1/2$ &
$3/2)^1$ & $-1/2$ & $-3/2$ \\[1mm] \hline
GAC & Asp D & $(1/2$ & $3/2)^2$ & $1/2$ & $-1/2$ & AAC & Asn N & $3/2$ &
$3/2$ & $-1/2$ & $-1/2$ \\
GAU & Asp D & $(1/2$ & $3/2)^2$ & $-1/2$ & $-1/2$ & AAU & Asn N & $3/2$ &
$3/2$ & $-3/2$ & $-1/2$ \\
GAG & Glu E & $(1/2$ & $3/2)^2$ & $1/2$ & $-3/2$ & AAG & Lys K & $3/2$ &
$3/2$ & $-1/2$ & $-3/2$ \\
GAA & Glu E & $(1/2$ & $3/2)^2$ & $-1/2$ & $-3/2$ & AAA & Lys K & $3/2$ &
$3/2$ & $-3/2$ & $-3/2$ \\[1mm] \hline
\end{tabular}
\end{center}
\end{table}

\clearpage

\begin{table}[htbp]
\caption{Data for vertebrates from GenBank [Release 149.0] 
\label{tabledata1}}
\footnotesize
\begin{center}
\begin{tabular}{|r|l|c|c|c|}
\hline
& Biological species & number of sequences & number of codons  & GC \% \\
\hline
1 & Bos taurus & 3526 & 1434728 & 53,81 \\
2 & Canis familiaris & 948 & 443570 & 53,29 \\
3 & Cavia  porcellus & 388 & 159307 & 52,07  \\
4 & Cricetulus griseus & 302 & 142938 & 51,02 \\
5 & Cyprinus carpio & 344 & 137658 & 49,89 \\
6 & Danio rerio & 12639 & 5184976 & 50,55  \\
7 & Equus caballus & 317 & 115073 & 52,92 \\
8 & Felis catus & 282 & 108779 & 52,54  \\
9 & Gallus gallus & 5498 & 2507341 & 51,21  \\
10 & Homo sapiens & 82409 & 35354305 & 52,41\\
11& Macaca fascicularis & 4029 & 1403004 & 49,17 \\
12 & Macaca mulatta & 683 & 224889 & 52,84 \\
13 & Mesocricetulus auratus  & 301 & 128917 & 52,56 \\
14 & Mus musculus & 39535 & 18330339 &  52,26 \\
15 & Mus sp. & 573  & 124703 &  52,77 \\
16 & Oncorhynchus mykiss & 725  & 271803 & 53,25 \\
17& Oryctolagus cuniculus & 1038  & 506266 & 54,7  \\
18 & Oryzias latipes & 434 & 193442 & 52,32  \\
19 & Ovis aries & 597 & 188842 & 53,48 \\
20 & Pan troglodytes & 272 & 281194 &  54,41 \\
21 & Rattus norvegicus & 13977 &  6477319 &  52,59 \\
22 & Rattus sp. & 654 & 127597 &  52,37 \\
23 & Sus scrofa & 1932 & 758646 & 54,25 \\
24 & Takifugu rubripes & 746 & 363990 & 54,61 \\
 25& Xenopus laevis & 10831 & 4760901 &  46,88 \\
26 & Xenopus tropicalis & 3080  & 1181338 & 47,8  \\
\hline
\end{tabular}
\end{center}
\end{table}


\newpage

\begin{table}[htbp]
\caption{Data for invertebrates from GenBank [Release 149.0]  
\label{tabledata2}}
\footnotesize
\begin{center}
\begin{tabular}{|r|l|c|c|c|}
\hline
& Biological species & number of sequences & number of codons  & GC \% \\
\hline
1 & Aedes aegyptiana & 454 & 198039 & 50,52 \\
2 & Anopheles gambiae & 578 & 248920 & 55,89 \\
3 & Bombyx mori & 751 & 300663 & 48,71  \\
4 &  Caenorhabditis elegans & 24353 & 10911983 & 42,93 \\
5 &  Ciona  intestinalis & 726 & 356806 & 45,18 \\
6 &  Cryptosporidium  parvum & 638 & 400196 & 33,28  \\
7 & Dictyostelium  discoideum & 3349 & 1953700 & 28,63 \\
8 &  Drosophila melanogaster & 38522 & 20352622 & 53,89  \\
9 & Drosophila pseudoobscura & 301 & 107621 & 54,85  \\
10 &Drosophila	simulans &	609	&	214015	&	52,59 \\
11&  Drosophila	subobscura  &		291	&	107270	&	55,64	 \\
12 & Drosophila	virilis &		214	&	114895	&	53,52	 \\
13 &  Encephalitozoon	cuniculi  GB-M11 &	1995 	& 718636	&	47,52	\\
14 &  Entamoeba	histolytica 6		285 &		115735 &		31,38 \\
15 &  Giardia	intestinalis &		355		 & 178606		& 52,02 \\
16 &  Leishmania	major  &		1714		 & 1050352	&	62,55	\\
17&  Manduca	sexta &		294	&	118722	&	50,49 \\
18 & Oikopleura	dioica &		387	&	185426	&	46,61	 \\
19 & Paramecium	tetraurelia &		617	&	354780	&	30,79 \\
20 & Plasmodium	falciparum &		1061	 &	652011	&	27,44 \\
21 &  Plasmodium	falciparum 3D7 &   4097 &		3031547	&	23,83\\
22 &  Plasmodium	vivax &		335	&	266303	&	43,03\\
23 & Schistosoma	mansoni &	303	&	124811	&	37,36 \\
24 & Strongylocentrotus	purpuratus &		239	&	144932	&	50,18 \\
 25& Tetrahymena	thermophila &		207	 &	108737	&	33,23 \\
26 & Toxoplasma	gondii &		433	&	213392	&	55,72 \\
27 &Trypanosoma	brucei &		5101	 &	2610083	 &	50,73  \\
28 &  Trypanosoma	cruzi &		702	&	304975	& 54,29	 \\
\hline
\end{tabular}
\end{center}
\end{table}

\begin{table}[ht]
\caption{Data for plants from GenBank [Release 149.0]
\label{tabledata3}}
\footnotesize
\begin{center}
\begin{tabular}{|r|l|c|c|c|}
\hline
& Biological species & number of sequences & number of codons  & GC \% \\
\hline
1 & Arabidopsis	thaliana & 	73134	&		28641535	 &	44,6 \\
2 & Ashbya	gossypii	ATCC	10895 &	4709	 & 2315585	&	52,57 \\
3 &  Aspergillus	fumigatus & 	639		&	331308	&	54,17\\
4 &  Aspergillus	niger &	229	&		111120	&	56,24	 \\
5 &  Aspergillus	oryzae &	238	&		127932	&	53,77	 \\
6 &  Botryotinia	fuckeliana &	127	&		120431	&	46,55 \\
7 & Brassica	napus &	519	&		190971	&	47,63 \\
8 & Candida	albicans &	691		&	390039	&	36,87  \\
9 & Candida	glabrata	CBS138 &	5165	 &	2607853	&	40,44 \\
10 & Chlamydomonas	reinhardtii & 	728	&		356299	&	66,23 \\
11  & Cochliobolus	heterostrophus &	106	&		150610		& 51,64  \\
12 & Cryptococcus	neoformans	var. &	6587	 &	3529040	&	51,17 \\
13 & Debaryomyces	hansenii	CBS767 & 6182	&	2865738	&	37,45  \\
14 & Emericella	nidulans &	640		&	386661	&	53,03 \\
15 & Glycine	max & 	957		&	395689	&	45,87 \\
16 & Gossypium	hirsutum & 	388		&	134962		& 45,82  \\
17 & Hordeum	vulgare & 	301	&		119908	&	55,1 \\
18 & Hordeum	vulgare	subsp.	vulgare &	1174 & 	337363	&	55,99 \\
19 & Lycopersicon	esculentum &	1249	 & 		543566	&	42,58  \\
20 & Medicago	truncatula &	285	&		122993	&	41,5 \\
21 & Neurospora	crassa &	3918	 &		2014863	&	56,13 \\
22 & Nicotiana	tabacum &	1325		&	506072	&	43,66 \\
23 & Oryza	sativa &	69548	&		24683258	 &	55,39  \\
24 & Phaseolus	vulgaris & 	259		&	110511	&	45,83 \\
25 & Physcomitrella	patens &	250	 &		112915	&	50,85 \\
26 & Pisum	sativum &	779	&		302738	&	43,24 \\
27 & Pneumocystis	carinii &	165		&	108663	&	35,45  \\
28 & Podospora	anserina &	247		&	123453	&	55,79  \\
29 & Saccharomyces	cerevisiae &	14164	&		6414021	&	39,75 \\
30 & Schizosaccharomyces	pombe &	6083		&	2849252	&	39,8  \\
31& Solanum	demissum & 	397		&	204262	&	41,04  \\
32 & Solanum	tuberosum &	741		&	319584	&	42,47  \\
33 & Sorghum	bicolor &	348	&		180635	&	54,36  \\
34 & Triticum	aestivum &	1221		&	446907		&  55,48  \\
35 & Ustilago	maydis &	175	&		113324	&	56,42 \\
36 & Yarrowia	lipolytica &	217	&		111091		& 54,6  \\
37 & Yarrowia	lipolytica	CLIB99 & 5967	 &	2945919		& 53,65 \\
48 & Zea	mays &	2194		&	930473	&	54,71 \\
\hline
\end{tabular}
\end{center}
\end{table}

\newpage

\begin{table}[hb]
\footnotesize
\begin{center}
\begin{tabular}{|r|l|c|c|c|}
\hline
& Biological species & number of sequences & number of codons  & GC \% \\
\hline
1 & Anolis	 allisoni &	96		& 33213		& 37,34 \\
2 & Anolis 	cybotes &	101	&	34921	&	35,34  \\
3 & Anolis	 sagrei &	316		& 108987		& 36,05  \\
4 & Bos	taurus &	789	&	228560	&	39,89  \\
5 & Canis	familiaris & 213		& 63324	&	39,62  \\
6 & Chaetodipus	intermedius &	231	&	60640	&	41,16  \\
7 & Gallus 	gallus &	103	&	30582	&	47,1  \\
8 & Heteronotia	binoei &	295	&	102365	&	44,94 \\
9 & Homo	sapiens &	17179	&	4913970	&	44,95  \\
10 & Microgale	longicaudata	& 104	&	36191	&	37,68  \\
11 & Microtus	oeconomus &	280	&	106679	&	44,07  \\
12 & Motacilla	alba & 172	&	59684	&	46,23 \\
13 & Mus	 musculus  &	167		& 45008	&	37,2  \\
14 & Nectarinia	olivacea   &	283	&	33109		& 47,22  \\
15 & Parus  	major & 	89	&	30912	&	49,25 \\
16 & Parus	 montanus  &	139	&	48231	&	50,24  \\
17 & Passerculus	sandwichensis  &	226	&	52252	&	47,04  \\
18 & Sus	scrofa &	478		& 150890	& 	40,3  \\
19 & Theragra	chalcogramma  & 	131	&	38607	&	41,97  \\
20 & Troglodytes	troglodytes  &	97	&	33659	&	38,87  \\
\hline
\end{tabular}
\caption{Data for mitochondrial  vertebrates.}  
\label{tabledata1m} 
\end{center}
\end{table}

\begin{table}[b]
\centering
\footnotesize \medskip
\begin{tabular}{|c||c|c|c|c|c|c|c|c|c|}
\hline
\begin{tabular}{c} Biological \\ species \end{tabular}
& $P_{C+A}(P)$ & $P_{C+A}(A)$ & $P_{C+A}(T)$ & $P_{C+A}(S)$ & $P_{C+A}(V) $
& $P_{C+A}(L)$ & $P_{C+A}(R)$ & $P_{C+A}(G)$ & $P'_{C+A}(S)$ \\
\hline
 1 & 0.60 & 0.63 & 0.64 & 0.60 & 0.35 & 0.33 & 0.51 & 0.59 & 0.65 \\
 2 & 0.59 & 0.61 & 0.65 & 0.59 & 0.36 & 0.34 & 0.52 & 0.59 & 0.65 \\
 3 & 0.59 & 0.62 & 0.65 & 0.60 & 0.37 & 0.34 & 0.52 & 0.59 & 0.65 \\
 4 & 0.60 & 0.59 & 0.62 & 0.59 & 0.35 & 0.31 & 0.53 & 0.58 & 0.67 \\
 5 & 0.60 & 0.60 & 0.62 & 0.56 & 0.38 & 0.33 & 0.50 & 0.58 & 0.62 \\
 6 & 0.60 & 0.63 & 0.65 & 0.60 & 0.35 & 0.33 & 0.52 & 0.60 & 0.65 \\
 7 & 0.53 & 0.56 & 0.61 & 0.56 & 0.34 & 0.32 & 0.55 & 0.63 & 0.63 \\
 8 & 0.61 & 0.65 & 0.64 & 0.63 & 0.35 & 0.33 & 0.55 & 0.61 & 0.66 \\
 9 & 0.60 & 0.63 & 0.63 & 0.60 & 0.37 & 0.35 & 0.50 & 0.58 & 0.65 \\
10 & 0.61 & 0.64 & 0.65 & 0.62 & 0.36 & 0.33 & 0.53 & 0.61 & 0.67 \\
11 & 0.61 & 0.62 & 0.64 & 0.60 & 0.38 & 0.34 & 0.52 & 0.59 & 0.65 \\
12 & 0.58 & 0.58 & 0.61 & 0.59 & 0.38 & 0.32 & 0.55 & 0.59 & 0.64 \\
13 & 0.61 & 0.63 & 0.66 & 0.60 & 0.35 & 0.35 & 0.55 & 0.61 & 0.67 \\
14 & 0.62 & 0.61 & 0.68 & 0.60 & 0.38 & 0.33 & 0.53 & 0.57 & 0.66 \\
15 & 0.61 & 0.62 & 0.66 & 0.58 & 0.37 & 0.33 & 0.51 & 0.59 & 0.64 \\
16 & 0.58 & 0.62 & 0.66 & 0.59 & 0.37 & 0.34 & 0.52 & 0.59 & 0.66 \\
17 & 0.62 & 0.54 & 0.72 & 0.59 & 0.29 & 0.35 & 0.58 & 0.58 & 0.69 \\
18 & 0.56 & 0.59 & 0.63 & 0.60 & 0.35 & 0.31 & 0.55 & 0.63 & 0.68 \\
19 & 0.61 & 0.61 & 0.65 & 0.61 & 0.34 & 0.34 & 0.50 & 0.59 & 0.64 \\
20 & 0.61 & 0.63 & 0.64 & 0.60 & 0.38 & 0.34 & 0.52 & 0.60 & 0.64 \\
21 & 0.58 & 0.63 & 0.66 & 0.62 & 0.37 & 0.34 & 0.53 & 0.61 & 0.68 \\
\hline
\end{tabular}
\caption{Sum of usage probability of codons $P_{C+A}(XN) \equiv
P(XNC)+P(XNA)$. The number in the first column denotes the biological
species of Table \ref{tabledata1}. The amino acid are labelled by the
standard letter. Morevover $P'_{C+A}(S) = P(UCA) + P(AGC)$.}
\label{tablePCA}
\end{table}

\clearpage

\begin{table}[ht]
\centering
\label{tablestatXZN}
\footnotesize \medskip
\begin{tabular}{|c||c|c|c|c|c|c|c|c|}
\hline
& $P(CCU)$ & $P(CCC)$ & $P(CCA)$ & $P(CCG)$ & $P(ACU)$ & $P(ACC)$ &
$P(ACA)$ & $P(ACG)$ \\
\hline
$\overline{x}$ & 0.28 & 0.33 & 0.26 & 0.13 & 0.23 & 0.39 & 0.26 & 0.13 \\
$\sigma$ & 0.028 & 0.043 & 0.034 & 0.028 & 0.030 & 0.050 & 0.034 & 0.027 \\
$\sigma/\overline{x}$ & 10.0 \% & 12.8 \% & 13.3 \% & 22.3 \% & 13.1 \% &
13.0 \% & 13.0 \% & 21.4 \% \\
\hline
\hline
& $P(GCU)$ & $P(GCC)$ & $P(GCA)$ & $P(GCG)$ & $P(UCU)$ & $P(UCC)$ &
$P(UCA)$ & $P(UCG)$ \\
\hline
$\overline{x}$ & 0.27 & 0.40 & 0.21 & 0.12 & 0.30 & 0.38 & 0.22 & 0.10 \\
$\sigma$ & 0.026 & 0.046 & 0.035 & 0.029 & 0.027 & 0.036 & 0.026 & 0.020 \\
$\sigma/\overline{x}$ & 9.5 \% & 11.6 \% & 16.5 \% & 25.3 \% & 8.8 \% & 9.3
\% & 12.0 \% & 20.2 \% \\
\hline
\hline
& $P(GUU)$ & $P(GUC)$ & $P(GUA)$ & $P(GUG)$ & $P(CUU)$ & $P(CUC)$ &
$P(CUA)$ & $P(CUG)$ \\
\hline
$\overline{x}$ & 0.17 & 0.26 & 0.10 & 0.47 & 0.15 & 0.25 & 0.08 & 0.52 \\
$\sigma$ & 0.036 & 0.023 & 0.026 & 0.045 & 0.034 & 0.018 & 0.017 & 0.035 \\
$\sigma/\overline{x}$ & 20.9 \% & 9.1 \% & 25.8 \% & 9.5 \% & 22.6 \% & 7.2
\% & 20.6 \% & 6.7 \% \\
\hline
\hline
& $P(CGU)$ & $P(CGC)$ & $P(CGA)$ & $P(CGG)$ & $P(GGU)$ & $P(GGC)$ &
$P(GGA)$ & $P(GGG)$ \\
\hline
$\overline{x}$ & 0.16 & 0.34 & 0.18 & 0.31 & 0.17 & 0.33 & 0.26 & 0.23 \\
$\sigma$ & 0.042 & 0.039 & 0.026 & 0.043 & 0.029 & 0.034 & 0.033 & 0.032 \\
$\sigma/\overline{x}$ & 26.0 \% & 11.4 \% & 14.0 \% & 13.9 \% & 16.9 \% &
10.3 \% & 12.7 \% & 13.7 \% \\
\hline
\end{tabular}
\caption{Mean value, standard deviation and their ratio for the
probabilities $P(XZN)$ corresponding to the eight amino-acids related to
quartets or sextets for the choice of biological species of Table \ref{tabledata1}.}
%
\centering
\label{tablestat-2}
\footnotesize \medskip
\begin{tabular}{|c||c|c|c|c|c|c|c|c|}
\hline
& $P_{C+A}(P)$ & $P_{C+A}(A)$ & $P_{C+A}(T)$ & $P_{C+A}(S)$ & $P_{C+A}(V) $
& $P_{C+A}(L)$ & $P_{C+A}(R)$ & $P_{C+A}(G)$ \\
\hline
$\overline{x}$ & 0.595 & 0.611 & 0.646 & 0.598 & 0.359 & 0.334 & 0.527 &
0.596 \\
$\sigma$ & 0.020 & 0.027 & 0.024 & 0.016 & 0.020 & 0.012 & 0.020 & 0.015 \\
$\sigma/\overline{x}$ & 3.4 \% & 4.4 \% & 3.8 \% & 2.6 \% & 5.6 \% & 3.7 \%
& 3.8 \% & 2.5 \% \\
\hline
\hline
& $P_{C+U}(P)$ & $P_{C+U}(A)$ & $P_{C+U}(T)$ & $P_{C+U}(S)$ & $P_{C+U}(V) $
& $P_{C+U}(L)$ & $P_{C+U}(R)$ & $P_{C+U}(G)$ \\
\hline
$\overline{x}$ & 0.613 & 0.672 & 0.614 & 0.687 & 0.430 & 0.401 & 0.506 &
0.507 \\
$\sigma$ & 0.030 & 0.028 & 0.027 & 0.026 & 0.031 & 0.022 & 0.049 & 0.024 \\
$\sigma/\overline{x}$ & 4.9 \% & 4.2 \% & 4.4 \% & 3.8 \% & 7.2 \% & 5.4 \%
& 9.7 \% & 4.8 \% \\
\hline
\hline
& $P_{C+G}(P)$ & $P_{C+G}(A)$ & $P_{C+G}(T)$ & $P_{C+G}(S)$ & $P_{C+G}(V) $
& $P_{C+G}(L)$ & $P_{C+G}(R)$ & $P_{C+G}(G)$ \\
\hline
$\overline{x}$ & 0.462 & 0.513 & 0.511 & 0.479 & 0.728 & 0.769 & 0.652 &
0.567 \\
$\sigma$ & 0.058 & 0.058 & 0.063 & 0.046 & 0.056 & 0.048 & 0.055 & 0.058 \\
$\sigma/\overline{x}$ & 12.4 \% & 11.3 \% & 12.3 \% & 9.6 \% & 7.7 \% & 6.3
\% & 8.4 \% & 10.2 \% \\
\hline
\end{tabular}
\caption{Mean value, standard deviation and their ratio for the sums of
probabilities $P_{C+A}$, $P_{C+U}$, $P_{C+G}$ corresponding to the eight
amino-acids related to quartets or sextets for the choice of biological
species of Table \ref{tabledata1}. The amino acid are labelled by the
standard letter.}
\end{table}

\newpage

\bibliography{Bibliografia}

\end{document}